\documentclass[aps, prd, superscriptaddress, nofootinbib, preprintnumbers, showpacs]{revtex4}
\usepackage{graphicx}

\newcommand{\beq}{\begin{equation}}
\newcommand{\eeq}{\end{equation}}
\newcommand{\beqs}{\begin{eqnarray}}
\newcommand{\eeqs}{\end{eqnarray}}
\newcommand{\zm}{z_{m}}
\newcommand{\ep}{\epsilon}
\newcommand{\tr}{\textrm{Tr}\,}
\newcommand{\condense}{\langle \bar{T}T \rangle}

\newcommand{\VEV}[1]{\langle #1 \rangle}

\begin{document}

 \title{{\bf   $S$ Parameter in the Holographic Walking/Conformal Technicolor
 \vspace{5mm}}}

 \author{{\bf Kazumoto Haba}}\thanks{
      {\tt haba@eken.phys.nagoya-u.ac.jp}}
       \affiliation{ Department of Physics, Nagoya University,
                    Nagoya, 464-8602, Japan.}
\author{{\bf Shinya Matsuzaki}}\thanks{
      {\tt synya@physics.unc.edu}}
      \affiliation{ Department of Physics and Astronomy,  
                    University of North Carolina, Chapel Hill 27599-3255.}
\author{{\bf Koichi Yamawaki}} \thanks{
      {\tt yamawaki@eken.phys.nagoya-u.ac.jp}}
      \affiliation{ Department of Physics, Nagoya University,
Nagoya, 464-8602, Japan.}

\begin{abstract}

We explicitly calculate the $S$ parameter in entire parameter space 
of the holographic walking/conformal technicolor (W/C TC), 
based on the deformation of the holographic QCD by varying the anomalous dimension from  $\gamma_m \simeq 0$ through $\gamma_m \simeq 1$ continuously. 
The  $S$ parameter is  given as a positive monotonic function 
of  $\xi$ which is fairly insensitive to $\gamma_m$ and 
continuously vanishes as $S \sim \xi^2 \to 0$ when  $\xi \to 0$,   
where  $\xi$ is the vacuum expectation value of the bulk scalar field at the infrared boundary of the 5th dimension 
$z=z_m$ and  is related to the mass of  (techni-) $\rho$ meson ($M_\rho$) and the decay constant ($f_\pi$) as 
$\xi \sim f_\pi z_m \sim f_\pi/M_\rho$ for $\xi \ll 1$. 
However, although $\xi$ is related to the techni-fermion condensate $\condense$, 
we find no particular suppression of $\xi$ and hence of $S$ due to large $\gamma_m$,  
based on the correct identification of the renormalization-point dependence 
of  $\condense$  in contrast to the literature. 
Then we argue 
possible behaviors of $f_\pi/M_\rho$ as $\condense \to 0$  
near the conformal window characterized by the Banks-Zaks infrared fixed point
 in more explicit dynamics with  $\gamma_m \simeq 1$. 
It is a curious coincidence that  the result from ladder Schwinger-Dyson and Bethe-Salpeter equations 
well fits in the parameter space obtained in this paper. When $f_\pi/M_\rho \to 0$ is realized, 
the holography suggests a novel possibility that $f_\pi$ vanishes much faster than 
the dynamical mass $m$ does.

\end{abstract}

\maketitle

\section{Introduction}

The origin of the electroweak symmetry breaking is the most urgent issue
to be resolved at the LHC experiments. 
In the standard model,  
Higgs boson is introduced 
just as a  phenomenological input 
only for the sake of making particles acquire masses. 
As such the standard model does not explain the origin of the electroweak symmetry breaking. 
With the existence of Higgs boson as an elementary particle, moreover, 
one necessarily faces with some problems such as naturalness, etc.

One of the candidates which resolve these problems is the technicolor (TC) model~\cite{Weinberg:1975gm}. 
In the framework of TC, 
the origin of the electroweak symmetry breaking is explained dynamically 
without 
introduction of Higgs boson. 
The simplest model of TC, 
just a simple scale-up 
of QCD, however, 
does not pass the electroweak precision test, especially, 
suffers from a large contribution of ${\cal O}(1)$ to the Peskin-Takeuchi  $S$ parameter~\cite{Peskin:1990zt+}, 
while the electroweak precision test shows that 
the value of the $S$ parameter is less than  
about 0.1.

  There is an interesting possibility~\cite{Appelquist:1991is,Harada:2005ru} that 
contributions to the $S$ parameter can be reduced 
in the case of the walking/conformal TC 
(W/C TC)~\cite{Holdom:1984sk,Yamawaki:1985zg, Akiba:1985rr, Appelquist:1986an},
initially dubbed as ``scale-invariant TC"~\cite{Yamawaki:1985zg}, 
with  almost non-running (walking) gauge coupling near the conformal fixed point, 
which produces a large anomalous dimension $\gamma_m \simeq 1$ of the techni-fermion condensate operator
$\bar T T$~\cite{Yamawaki:1985zg} (for reviews, see \cite{Miransky:vk}). 
A salient feature of this theory is the appearance of  a composite Higgs boson as a 
 (massive) techni-dilaton~\cite{Yamawaki:1985zg} associated with the 
spontaneous 
breaking of the scale invariance (as well as the explicit breaking due to the scale anomaly) . 
A typical example~\cite{Appelquist:1996dq, Miransky:1996pd}  of such a W/C TC is 
based on the Banks-Zaks infrared fixed point~\cite{Banks:1981nn} (BZ-IRFP)  $\alpha_* $ 
in the large $N_f$ QCD, QCD with many massless flavors $N_f \gg 3$ ($N_f < 11N_c/2)$).
Looking at the region $0< \alpha < \alpha_*$, we note that  $ \alpha_* \searrow 0$ when $N_f  \nearrow 11 N_c/2$, 
and hence there exists a certain region $(N_f^* <) \, N_f^{\rm cr}< N_f < 11 N_c/2 $ (``conformal window'') 
such that   $ \alpha_*  < \alpha^{\rm crit}$,  
where $\alpha^{\rm crit} $ is the critical coupling for the spontaneous chiral symmetry breaking 
and hence the chiral symmetry gets restored in this region.
Here $\alpha^{\rm crit} $ may be evaluated as 
$\alpha^{\rm crit} =\pi/3 C_2(F)$ in the ladder approximation, 
in which case  we have $N_f^{\rm cr} \simeq 4 N_c$~\cite{Appelquist:1996dq} 
\footnote{
In the case of  $N_c=3$, this value $N_f^{\rm cr} \simeq 4 N_c =12$ is somewhat different 
from the lattice value~\cite{Iwasaki:2003de} \ $6<N_f^{\rm cr}<7$, but is consistent with more recent lattice results~\cite{Appelquist:2007hu}.
} \
\footnote{
There is another possibility for the W/C TC with much less $N_f$ based on the higher TC representation~\cite{Hong:2004td},
although explicit ETC model building would be somewhat involved.}. 
Related to the conformal symmetry, this phase transition (``conformal phase transition"\cite{Miransky:1996pd}) has 
unusual nature that the order parameter changes continuously but the spectrum does discontinuously 
at the phase transition point.

When it is applied to TC, we set $\alpha_*$ slightly larger than $\alpha^{\rm crit}$ 
(slightly outside of the conformal window),  with the running coupling becoming
 larger than the critical coupling only in the infrared region, 
we have a condensate or the dynamical mass of the techni-fermion $m$ 
of the order of such an  infrared scale which is much smaller than
the intrinsic scale of the theory $\Lambda_{\rm TC} (\gg m )$. 
Although the  BZ-IRFP actually disappears due to decoupling of  massive fermion at the scale of $m$,  the coupling is 
still walking due to the remnant of the BZ-IRFP 
in a wide region $m < \mu < \Lambda_{\rm TC} $. 
Then the theory develops a large anomalous dimension $\gamma_m \simeq 1$ 
and enhanced condensate $\VEV{\bar T T}|_{\Lambda_{\rm TC}} 
\sim \Lambda_{\rm TC}  m ^2$ at the scale of $\Lambda_{\rm TC}$ 
which is usually identified with the ETC scale $\Lambda_{\rm TC} =\Lambda_{\rm ETC}$~\cite{Appelquist:1996dq, Miransky:1996pd}. 
Note that   $m\to 0$ as $\alpha_*  \searrow\alpha^{\rm crit}$  and the mass of techni-dilaton, $M_{\rm TD} \simeq 
\sqrt{2} m$,~\footnote{
This estimate~\cite{Shuto:1989te}  is based on the  ladder Schwinger-Dyson (SD) equation for the gauged Nambu-Jona-Lasinio model which well simulates~\cite{Appelquist:1996dq,Miransky:1996pd} the 
conformal phase transition in the large $N_f$ QCD. Actually, the result 
is consistent with the straightforward calculation~\cite{Harada:2003dc} of scalar bound state mass,
$M_{\rm TD} \sim 1.5\, m$, through  coupled use of the SD equation and 
(homogeneous) Bethe-Salpeter (BS) equation in the ladder approximation.}
 also vanishes to be degenerate with the Nambu-Goldstone (NG) boson, 
although there is no light spectrum in the conformal window as a characteristic feature of the conformal phase transition.

The W/C TC,  
however, has a calculability problem, since its 
non-perturbative dynamics is not QCD-like at all, 
and hence no simple scaling of QCD results would be available. 
The best thing we could do so far has been a straightforward calculation 
based on the SD equation and (inhomogeneous) BS equation 
in the ladder approximation~\cite{Harada:2005ru}, 
which is however not a systematic approximation and is not very reliable in the quantitative sense.

   Of a late fashion, based on the so-called AdS/CFT correspondence,  
a duality of the string in the anti-de Sitter space background-conformal field theory~\cite{Maldacena:1997re}, 
holography gives us a new method which may resolve 
the calculability problem of strongly coupled gauge theories~\cite{Arkani-Hamed:2000ds}: 
Use of the holographic correspondence enables us to 
calculate Green functions in a four-dimensional strongly coupled theory 
from a five-dimensional weakly coupled theory. 
For instance, QCD can be reformulated based on the holographic correspondence 
either in the bottom-up approach~\cite{DaRold:2005zs,Erlich:2005qh}  
or in the top-down approach~\cite{Sakai:2004cn}. 
In both approaches we end up with the five-dimensional gauge theory for the flavor symmetry, 
whose infinite tower of Kaluza-Klein modes describe nicely a set of the massive vector/axialvector mesons 
as the gauge bosons of Hidden Local Symmetries (HLS)~\cite{Bando:1984ej,Bando:1985rf,Harada:2003jx}, 
or equivalently as the Moose~\cite{Georgi:1985hf}. 
Although a holographic description is valid only for large $N_c$ limit, 
several observables of QCD have been reproduced 
within 30~\% errors in both approaches. Moreover,  
through the high-energy behavior of current correlators  in operator product expansion, 
some consistency with the QCD
has been confirmed in the bottom-up approach.

Recently several authors~\cite{Hong:2006si,Piai:2006hy}  calculated 
the $S$ parameter in the  W/C TC as an application of the above technique of 
bottom-up holographic QCD~\cite{DaRold:2005zs,Erlich:2005qh} 
to the holographic W/C TC: They made  some deformation 
adjusting a profile of a 5-dimensional bulk scalar field 
which is related to the anomalous dimension of techni-fermion condensate $\gamma_m$.  
They claimed  that when $\gamma_m \equiv 1$, 
the  $S$ parameter  for certain parameter choices 
 is substantially  reduced compared to that of the  QCD-like theory 
with $\gamma_m \simeq 0$. 
It is not clear, however, how the non-trivial feature of the dynamics 
of walking/conformal theory contributes to that reduction, 
since they discuss only specific parameter choices relevant to specific TC models.
Actually,  {\it it is not $\gamma_m \equiv 1$ but  $\gamma_m \simeq 1$} ($\gamma_m <1$) that  is needed for 
realistic model building of  W/C TC where $\gamma_m=1$ should be regarded an idealized limit of  $\gamma_m \to
1$ from the side of $\gamma_m <1$.

   In this paper, based on the holographic correspondence in the bottom-up approach, 
we calculate the $S$ parameter in the W/C TC, 
treating the anomalous dimension $\gamma_m$ 
as a free parameter as $0<\gamma_m <1$, varying {\it continuously} from the QCD monitor value  $\gamma_m \simeq 0$ 
through the one of the W/C TC  $\gamma_m \simeq 1$.
We  calculate $S$ 
as  an explicit  function of  $\xi \equiv L v(z) \Big|_{z=z_m}$  
in the entire region of  $\xi$, 
where $L$ is the ${\rm AdS}_5$ radius, 
$v(z) \equiv \langle \Phi(x^\mu,z)\rangle$ is the vacuum expectation value of the bulk scalar field 
$\Phi(x^\mu,z)\,\, (\mu=0,1,2,3)$ 
and the 5th dimension $z$ has both infrared cutoff $z_m$ and ultraviolet cutoff $\epsilon$:  $\epsilon <z < z_m$.
This is in contrast to the previous authors~\cite{Hong:2006si,Piai:2006hy} 
whose discussions correspond to specific values of the parameter $\xi$  
and are restricted to the case of $\gamma_m \equiv 1$ as the W/C TC. 
Since the realistic model building of  W/C TC is {\it not} for
$\gamma_m \equiv 1$ but  $\gamma_m \simeq 1$ ($(\gamma_m <1$) , the analysis of  Ref.~\cite{Hong:2006si,Piai:2006hy} 
could be  a too much idealization, unless their  result  is continuously connected with the limit $\gamma_m \to 1$ from the side of  $0< \gamma_m <1$.   Actually, it turns out that 
the analysis of Ref.~\cite{Hong:2006si} is {\it not} continuously connected with the
$\gamma_m \to 1$ limit of our result and thus would not precisely correspond to the realistic situation of the W/C TC
we are interested in.

Then we find that 
$S$ is a positive function of  $\xi$ 
in accord with the previous authors~\cite{Hong:2006si, Agashe:2007mc}. 
We also find that both $z_m f_\pi$ and $ S$ are monotonically increasing functions of $\xi$ which is related to 
chiral condensate $\condense$. Noting that $z_m^{-1} \sim M_\rho$  
with $M_\rho$ being the techni-$\rho$ meson mass, we have an expression of $S$ as a function of $f_\pi/M_\rho$.  

 Most remarkably, we find that 
$S$ {\it continuously goes to zero} as $S \sim \xi^2N_{\rm TC} \sim  (f_\pi/M_\rho)^2 \to 0$ 
if  $ \xi  \to 0$ (See Fig.\,\ref{S-compare}), which is also in accord with the previous author~\cite{Piai:2006hy} 
discussing the case of $\gamma_m \equiv  1$. 
We also find that the result is  fairly insensitive to the value of  $\gamma_m$ unless we have substantial reduction of
$\xi$ or $f_\pi/M_\rho$ for larger  $\gamma_m$: 
Writing $\hat S=S/(N_f/2) =B  f_\pi^2/M_\rho^2$, we find  $B \simeq 27$ for $\gamma_m \simeq 1 $ and  $B \simeq 32$
for $\gamma_m \simeq 0 $  (See Fig.\,\ref{FigforB}). 
Although the result  of  (slight) decreasing tendency for fixed  $f_\pi/M_\rho$ is not inconsistent with Ref. \cite{Hong:2006si},  their value for  $\gamma_m \equiv 1$ is  somewhat smaller than ours with $\gamma_m \to 1$.

Our result roughly coincides with the well-known fact (see e.g. Ref.~\cite{Harada:2003jx}) 
that $\hat S$  in QCD case is given by the $\rho$ meson dominance
as  $\hat S=- 16 \pi \, L_{10} =4 \pi \cdot (F_\rho/M_\rho)^2 = 4\pi a  (f_\pi/M_\rho)^2=(g^2/(4\pi))^{-1}$,  with $a\simeq 2$ experimentally,
where $F_\rho=\sqrt{a} f_\pi $ is the decay constant of the $\rho$ meson (or, of the fictitious Nambu-Goldstone  boson absorbed into the 
$\rho$ meson in the language of HLS)  and $g$ the gauge coupling of HLS. However, our result  is highly nontrivial,  since the holography includes an infinite tower of  the vector and axialvector meson poles not just the lowest $\rho$ pole contribution.
Note that   the contributions of vector mesons are opposite in sign to those of  the axialvector mesons 
 and therefore the infinite sum of all contributions could in principle result in any functional form such as giving a 
non-vanishing constant in the limit  $f_\pi/M_\rho  \to 0$.

It thus opens a novel possibility for having small $S$ parameter, if we find a mechanism of 
suppressing $\xi^2N_{\rm TC}  \sim f_\pi/M_\rho \ll 1 $
particularly near the conformal phase transition  $m \to 0$.
Then the next issue is whether or not we can realize $ \xi^2 N_{\rm TC}\sim f_\pi/M_\rho \ll 1 
$ in the W/C TC.

It is also to be noted that the above continuous vanishing of $S$ in the case that  $\xi^2N_{\rm TC} 
 \sim (f_\pi/M_\rho)^2 \to 0$ 
is highly nontrivial 
in sharp contrast to the usual perturbative calculation where 
the $S$ parameter does not vanish even in the chiral restoration limit 
$m \to 0$, i.e., 
$S \to \frac{N_f}{2} \frac{N_{\rm TC}}{6\pi}$ for $N_{\rm TC}$ technicolors and $N_f$ techni-flavors, 
although it is identically zero when $m=0$ as it should, i.e. 
there is a discontinuity at the chiral phase transition.

Unfortunately, the holographic approach  as it stands cannot decide 
whether or not $ \xi^2 N_{\rm TC}\sim (f_\pi/M_\rho)^2 \to 0$: 
 $f_\pi/M_\rho$ is given by a certain function of $\xi$ and $\gamma_m$ 
 which are both arbitrary parameters in this approach. 
Although $\xi$ is related with the chiral condensate $\condense$ which vanishes at the conformal phase transition point, 
we find no direct suppression of $\xi$ or $f_\pi/M_\rho$  and hence of $S$ due to the large $\gamma_m$
in contrast to the previous authors~\cite{Hong:2006si,Piai:2006hy}. 
Based on the correct identification of
the renormalization scale of $\condense$, we have  $\xi \sim (m z_m)^{3-\gamma_m}\sim (m/M_\rho)^{3-\gamma_m}$,
independently of  the non-physical renormalization point $L$ or $\epsilon$, 
which may or may not be small even if  $\condense \to 0$, unless we know  $m/M_\rho \ll 1$. Thus
$\xi$ is not necessarily a small parameter in this framework
even for $m \to 0$, not to mention for $L \to 0$, $\epsilon \to 0$.
Then the only possibility  to realize   $\xi^2 N_{\rm TC} \sim (f_\pi/M_\rho)^2 \to 0$ 
would be to discuss more concrete dynamics approaching
 the conformal phase transition where 
we have $m \to 0$ 
(or $f_\pi \to 0$) not just a large anomalous dimension $\gamma_m \simeq 1$. Actually the effects of 
anomalous dimension are highly involved, combined with the  
scaling of $f_\pi/M_\rho$ as  $m \to 0$,  as seen in the direct calculation  
based on the ladder SD and BS equations~\cite{Harada:2005ru}.

We then discuss possible scaling behavior of $\xi^2 N_{\rm TC} \sim (f_\pi/M_\rho)^2 \to 0$ 
near the conformal window  
  $m \rightarrow 0$. Although a  simple large $N_c$ argument  would  always imply
$\hat S \sim  \xi^2 N_{\rm TC} \sim (f_\pi/M_\rho)^2 \sim   N_{\rm TC}  \to \infty$, the conformal phase transition takes place due to
the Banks-Zaks infrared fixed point which only can be realized for large $N_f$ with $N_c/N_f =$ fixed. Then the behavior of  
  $ (f_\pi/M_\rho)^2$ near the conformal phase transition is highly nontrivial.
Obviously three possibilities in the limit of $m \to 0$: 
  i) $f_\pi/M_\rho \to \infty$,  
  ii) $f_\pi/M_\rho \to {\rm constant} \ne 0$, 
  iii) $f_\pi/M_\rho \to 0$. 

We find that the case i) 
is realized only for $\xi  \gg 1$, 
since $f_\pi/M_\rho$ is the monotonically increasing function of $\xi  $.
In this case we have $\frac{f_\pi}{\sqrt{N_{\rm TC}}} \sim m$,
which is the familiar scaling relation realized in QCD. 
Actually the case i) corresponds to the Vector Manifestation proposed 
in the HLS loop calculation~\cite{Harada:2003jx,Harada:2000kb}.

The case ii) where $f_\pi/M_\rho \to {\rm constant} \ne 0$
is realized only for the case 
$\xi \sim (m/M_\rho)^{3-\gamma_m} \to {\rm {constant}} \neq 0$ 
and hence $S \to $ constant $\ne 0$  for $m \to 0$.
In this case $\frac{f_\pi}{\sqrt{N_{\rm TC}}}  \sim M_\rho \sim m$, which is the same scaling relation as the case i).
The case ii) actually corresponds to the straightforward calculation 
based on the ladder SD and BS equations~\cite{Harada:2003dc, Harada:2005ru}.  It is amusing that a set of  
 $(\xi, \hat S/N_{\rm TC})$,  $\xi$ obtained from 
 homogeneous BS equation  and   $\hat S/N_{\rm TC}$ from
inhomogeneous BS equation both combined with  SD equation, 
well coincides with a single point on the line of  the   $(\xi,\hat S/N_{\rm TC})$-plane obtained in this paper.

The most interesting case for the TC is case iii) 
in which we have $ f_\pi/M_\rho \to 0 $ as $m \to 0$. 
We find that the case iii) 
is realized only for $\xi \ll 1$, 
since $f_\pi/M_\rho$ is a monotonically increasing function of $\xi$, although
we have no explicit dynamics at this moment.  We shall discuss some possible dynamics for this case which will
be tested by future studies.
In the case iii)  we find a novel scaling property of $f_\pi$ vanishing much faster than 
$m$ near the conformal window, resulting in the
 form $\frac{f_\pi}{\sqrt{N_{\rm TC}}} \sim m (m/M_\rho)^{2-\gamma_m}$,  which is quite different from the familiar one 
$\frac{f_\pi}{\sqrt{N_{\rm TC}}} \sim m $.
This could be testable by lattice calculation for large $N_f$ QCD. 

Although the bottom-up approach of the holography does not explicitly uses the large  $N_{\rm TC}$,  the top-down
approach needs that limit.  Then the result here might be potentially valid  only for large $N_{\rm TC}$ not near the
conformal phase transition region where $N_f$ is large with $N_f/N_{\rm TC}=$ fixed.
Nevertheless, the result of this paper 
$\hat S \sim (f_\pi/M_\rho)^2$ might be valid beyond the leading order of  $1/N_{\rm TC}$.
Then it  would  be highly desired to investigate the possibility for $f_\pi/M_\rho \to 0$ in some explicit dynamics.

The paper is organized as follows:

In Sec.~\ref{review}
we briefly review the framework of 
calculations in the holographic W/C TC 
model of Refs.~\cite{Hong:2006si,Piai:2006hy} based on the bottom-up holographic QCD~\cite{DaRold:2005zs,Erlich:2005qh}.

In Sec.~\ref{S-parameter}
we calculate the $S$ parameter in models holographically dual to  
W/C TC allowing for varying values of the 
large anomalous dimension of techni-fermion condensation 
$\gamma_m$ from the QCD monitor value $\gamma_m \simeq 0$ to the W/C TC value $\gamma_m \simeq 1$.

In Sec.~\ref{RGpoint}  we identify the renormalization-point of the $\condense$, based on which we find 
that there is no suppression factor solely due to large $\gamma_m$.  

In Sec.~\ref{critical}
 we classify holographic W/C TC models  
into three cases,  i) $f_\pi/M_\rho \to \infty$, ii) $f_\pi/M_\rho \to $ constant $\ne 0$, iii) $f_\pi/M_\rho \to 0$
as $m \to 0$ near the conformal window arising due to the Banks-Zaks infrared fixed point 
with $\gamma_m \simeq 1$. 
As an explicit dynamics for the case ii) we find a curious coincidence of the result of  
ladder SD and BS equations 
with the result in this paper.
It is also shown that if the case iii) is realized,  
the $S$ parameter goes to zero at the edge of the conformal window in such a way that 
$f_\pi \to 0 $ scales as $m \to 0$ much faster than the familiar form, $f_{\pi}^2 \sim m^2$.

Sec.~\ref{summary} is devoted to summary and discussion. 

In Appendix \ref{cond:gamma1} we discuss subtlety of the limit $\gamma_m \to 1$ and $\gamma_m =1$.

In Appendix \ref{PS}   we discuss the Pagels-Stokar formula in comparison with the holographic result.

\section{Review of Holographic Calculations} 
\label{review}

In this section 
we briefly review the framework of 
calculations in the holographic W/C TC 
model of Refs.~\cite{Hong:2006si,Piai:2006hy} with $\gamma_m = 1$
which is the deformation of the the bottom-up holographic QCD~\cite{DaRold:2005zs,Erlich:2005qh} 
with $\gamma_m = 0$
by adjusting a profile of a 5-dimensional bulk scalar field. 
 Here we consider a generic case with $0 < \gamma_m < 1$.

A holographic model~\cite{DaRold:2005zs,Erlich:2005qh,Hong:2006si,Piai:2006hy} 
is defined on the 5-dimensional anti-de Sitter space (AdS$_5$) 
with the metric, 
\begin{equation} 
ds^2= g_{MN} dx^M dx^N 
= 
\left(\frac{L}{z}\right)^2\big(\eta_{\mu\nu}dx^\mu dx^\nu-dz^2\big)
\,, \label{AdSmetric}
\end{equation}
where $\eta_{\mu\nu}={\rm diag}[1, -1, -1,-1]$ is a metric 
on 4-dimensional space-time spanned by the coordinate 
$x_\mu$, and $L$ denotes the curvature radius of AdS$_5$. 
The fifth direction $z$ is compactified on the interval, 
\begin{equation} 
\epsilon \leq z \leq z_m 
\,. 
\end{equation}

A holographic action~\cite{DaRold:2005zs,Erlich:2005qh} 
possessing an 
$SU(N_f)_L\,\times\,SU(N_f)_R$ gauge symmetry in 
5 dimensions is constructed from 
$SU(N_f)_{L,R}$ gauge fields $L_M(x,z)$ and $R_M(x,z)$, 
and a scalar field $\Phi(x,z)$ transforming  
under the $SU(N_f)_L\,\times\,SU(N_f)_R$ gauge symmetry 
as a bi-fundamental representation. 
The action is given by
\footnote{ Here ${\rm Tr}[T^a T^b]=\frac{1}{2}\delta^{ab}$
 and $L(R)_{MN} = \partial_M L(R)_N - \partial_N L(R)_M 
 - i [ L(R)_M, L(R)_N ]$. }, 
\begin{eqnarray}
S_5 &=& 
\,\frac{1}{g_5^2} \int\,d^4 x\,\int_{\ep}^{\zm}\,d\,z~\sqrt{g}\, 
\nonumber \\
&&\times \left(
-\frac{1}{2} \tr\left[{L_{MN}L^{MN}}+{R_{MN}R^{MN}}\right]+\tr\left[{D_M\Phi^\dagger D^M\Phi}
-m^2_5\Phi^\dagger \Phi \right]\right)
\,, 
\label{S5}
\end{eqnarray}
where $g_5$ denotes the gauge coupling in 5 dimensions 
and 
$g={\rm det}[g_{MN}]=(L/z)^{10} $. 
The covariant derivative acting on the scalar field $\Phi$ is defined as 
\begin{equation}
D_M\Phi=\partial_M \Phi+iL_M\Phi-i\Phi R_M
\, . 
\end{equation} 
This $\Phi$ may be parametrized 
by using scalar and pseudo-scalar fields, $\phi$ and $P$, 
as 
\begin{equation}
\Phi(x,z) = \phi(x,z) \exp[{iP(x,z)/v(z)}]
\,, 
\end{equation}
with $v(z)=\frac{1}{\sqrt{2}} \langle \phi \rangle$ being the vacuum expectation value (VEV) 
of $\Phi$.

 For later convenience, we introduce 
5-dimensional vector and axialvector gauge fields 
$V_M$ and $A_M$ defined by 
\begin{equation}
V_M=\sqrt{\frac{1}{2}} \big(L_M+R_M) 
\,, \qquad 
A_M=\sqrt{ \frac{1}{2} }\big(L_M-R_M) 
\, ,
\end{equation}
and we choose a gauge, 
\begin{equation} 
 V_z(x,z) = A_z(x,z) \equiv 0 
\,.
\label{gauge-fixing} 
\end{equation}

 Based on AdS/CFT correspondence, 
boundary conditions for the bulk fields $V_M$, $A_M$, and $\Phi$ 
are chosen so that their UV boundary values are related to 
the external sources in TC theories in the limit of $\epsilon \to 0$:  
For the VEV of $\Phi$, 
the UV boundary value is related to 
the external source for 
the techni-fermion condensate $\condense$, namely, 
the current mass of techni-fermion $M$
in such a way that 
\begin{eqnarray}
M &\equiv& \lim_{\epsilon \to 0} {\cal{M}}, \nonumber \\
{\cal {M}} &=& 
\left(\frac{L}{\epsilon}\right)^{\gamma_m} \left(\frac{L}{\epsilon} v(z)\right)
\,\Big|_{z=\epsilon}, 
\label{def:M}
\end{eqnarray} 
where $\gamma_m$ stands for 
the anomalous dimension of the techni-fermion condensate 
$\langle \bar{T}T  \rangle$.
The AdS/CFT correspondence 
makes it possible to associate $m_5$, 
the mass of the scalar field $\Phi$, 
with the anomalous dimension $\gamma_m$: 
\begin{equation}
m_5^2\,=\,- \frac{(3-\gamma_m)(\gamma_m+1)}{L^2}
\,. 
\end{equation} 
We introduce the variable $\xi$ for the IR boundary value of VEV of $\Phi$,
\begin{eqnarray}
\xi = L v(z) \,\Big|_{z=z_m} \, ,
\label{vbdry}
\end{eqnarray} 
which corresponds to 
$\langle \bar{T}T  \rangle$ as will be seen later (Eq.(\ref{condenseL})). 

We shall later discuss $M$ and the corresponding $\langle \bar{T}T \rangle$ 
are quantities renormalized at the scale $1/L$ (see Sec.\ref{fpi:mscaling}),
whereas $\xi$ is the quantity renormalized at $1/z_m$.

As for the bulk gauge fields $V_\mu$ and $A_\mu$,  
the UV boundary values 
play the role of the external sources ($v_\mu$, $a_\mu$) 
for the vector and axialvector currents 
coupled to the holographic TC. 
Accordingly, under $V_z \equiv A_z \equiv 0$ gauge, the boundary condition may be chosen, 
\begin{eqnarray}\label{VAmuBC}
&& \partial_z V_{\mu}(x,z)\big|_{z=z_m} =
\partial_z A_{\mu}(x,z)\big|_{z=z_m}=0 
\,, \nonumber \\ 
&&
V_{\mu}(x,z)\big|_{z=\epsilon}=v_{\mu}(x) 
\,, \qquad 
A_{\mu}(x,z)\big|_{z=\epsilon}=a_{\mu}(x) 
\, .
\end{eqnarray}

 With these boundary conditions (\ref{vbdry}) and (\ref{VAmuBC}), 
the equations of motion for the bulk gauge fields are completely solved 
at the classical level. 
By substituting those solutions into the action (\ref{S5}), 
the effective action $W$ is expressed as a functional of the 
UV boundary values/external sources, ${\cal {M}}$, $v_\mu$, and $a_\mu$, i.e., 
$ W = W[{\cal {M}}, v_\mu, a_\mu]$. 
The two-point Green functions are then readily calculated as  
\begin{eqnarray} 
\frac{ \delta^2 W[v_\mu]}{\delta \tilde v^a_\mu(q) \delta  \tilde v^b_\nu(0)} \Bigg|_{v_\mu=0} 
&=& 
i\int_x \,e^{iq\cdot x} \langle 
J^{a\mu}_V (x) J^{b\nu}_V(0) \rangle = 
-\delta^{ab} \left(g^{\mu\nu}-\frac{q_{\mu}q_{\nu}}{q^2}\right)\,  \Pi_{V}(-q^2) 
\, , \nonumber \\ 
\label{Pi:V} \\ 
\frac{ \delta^2 W[a_\mu]}{\delta \tilde a^a_\mu(q)  \delta \tilde a^b_\nu(0)} \Bigg|_{a_\mu=0} 
&=& 
i\int_x \,e^{iq\cdot x} \langle 
J^{a\mu}_A (x) J^{b\nu}_A(0) \rangle = 
-\delta^{ab} \left(g^{\mu\nu}-\frac{q_{\mu}q_{\nu}}{q^2}\right)\, \Pi_{A}(-q^2) 
\, ,\nonumber \\ 
\label{Pi:A} \\ 
\lim_{\ep \to 0} i \frac{ i \delta W[{\cal {M}}]}{\delta {\cal {M}} } \Bigg|_{{\cal {M}}=0} 
&\equiv &\lim_{\ep \to 0}  i \frac{ i \delta W[M]}{\delta M } \Bigg|_{M=0} 
=
\langle \bar{T} T \rangle 
\,, \label{TbarT}
\end{eqnarray}
where $\tilde{v}^\mu(q)$ and $\tilde{a}^\mu(q)$ respectively 
denote  the Fourier component of $v^\mu(x)$ and $a^\mu(x)$. 

Once the current correlators are calculated, we can compute the $S$ parameter.
We define $\hat{S}$ as the $S$ parameter per each techni-fermion doublet, 
\begin{equation} 
\hat{S} = \frac{S}{N_f/2 } 
\,, 
\end{equation}
which is expressed by the vector and axialvector current correlators $\Pi_V$ and $\Pi_A$ 
as 
\begin{equation}
\hat{S}=-4\pi\,\frac{d}{ d
Q^2}\left[\Pi_V(Q^2)-\Pi_A(Q^2)\right]_{Q^2=0}
\,, 
\label{Sdif}
\end{equation}
where $Q \equiv \sqrt {- q^2}$. 
 In the next subsections we shall calculate those current correlators, $\condense$, $\Pi_V$ and $\Pi_A$.

\subsection{Generating Functional $W[{\cal {M}}]$ and $\condense$}

   Let us focus on a portion of the action (\ref{S5}) relevant for the VEV of $\Phi(x,z)$, 
$\langle \phi \rangle = v(z)$:
\begin{equation}\label{vaction}
S_5\mid_{v}=\int d^4 x \int_\epsilon^{z_m} dz \, \frac{L^3}{2g_5^2} 
\tr \left[-\frac{1}{z^3} (\partial_{z}v(z))^2 
+ \frac{(3-\gamma_m)(1+\gamma_m)}{z^5} v^2(z) \right] 
\, ,
\end{equation}
which leads to 
the following classical equation of motion for $v(z)$: 
\begin{equation}\label{vEOM}
\partial_{z}\left(\frac{1}{z^3}\partial_{z}v(z) \right) + 
\frac{(3-\gamma_m)(1+\gamma_m)}{z^5}v(z)=0 
\, . 
\end{equation}    
 Solution for $0 < \gamma_m <1 $ is given by
\footnote{
 If we set $\gamma_m \equiv 1$ in Eq.(\ref{vEOM}), 
we find a solution $v(z)=C_1 z^2 + C_2 z^2 \ln{z}$, which was used in analysis in Refs.\cite{Hong:2006si, Piai:2006hy}. 
Here we understand $\gamma_m = 1$ as the limit $\gamma_m \to 1 - 0$ which implies $C_2 \to 0$~\cite{Piai:2006hy}, namely,
$v(z) = \Sigma \left(\frac{z}{L}\right)^{2} = \left(\frac{L}{z_m}\right)^{2}\frac{\xi}{L}\left(\frac{z}{L}\right)^{2}$ 
as seen from Eq.(\ref{Sigma:ch}).
The other choice $C_1 = 0$ was adopted in Ref.\cite{Hong:2006si}. 
See Appendix \ref{cond:gamma1} for 
discussion on this point.
\label{foot:coeff}} 
\begin{equation}
v(z)^{(\epsilon)} =c_1 \left(\frac{z}{L}\right)^{1+\gamma_m}+ c_2\left(\frac{z}{L}\right)^{3-\gamma_m} 
\,, 
\end{equation} 
where $c_1$ and $c_2$  are determined by the boundary conditions (\ref{def:M}) and (\ref{vbdry}) as 
\begin{eqnarray}
\label{vcoffi1}
c_1 &=& \frac{{\cal {M}}-\left(\frac{\epsilon}{z_m}\right)^{2-2\gamma_m} \left(\frac{L}{z_m}\right)^{1+\gamma_m}\frac{\xi}{L} }
{\left( 1-\left(\frac{\epsilon}{z_m} \right)^{2-2\gamma_m}\right)}
\,,\\ 
\label{vcoffi2}
c_2 &=& \frac{1}{L}\frac{\left(\frac{L}{z_m}\right)^{3-\gamma_m}\xi-(\frac{L}{z_m})^{2-2\gamma_m} L {\cal {M}}}
{\left( 1-\left(\frac{\epsilon}{z_m} \right)^{2-2\gamma_m}\right)} \nonumber \\
&=& \left(\frac{L}{z_m}\right)^{3-\gamma_m}\frac{\xi}{L}
- \left(\frac{L}{z_m}\right)^{2-2\gamma_m} c_1.
\end{eqnarray}
In the continuum limit $\epsilon \to 0$ the solution takes the  form
\begin{equation}\label{vsol}
v(z) =M \left(\frac{z}{L}\right)^{1+\gamma_m}+ \Sigma \left(\frac{z}{L}\right)^{3-\gamma_m} 
\,, 
\end{equation} 
where  $\lim_{\epsilon \to 0} c_1 =\lim_{\epsilon \to 0} {\cal M} =M $ and 
 $\lim_{\epsilon \to 0} c_2 = \Sigma $ are quantities renormalized at the scale $1/L$
 and we may write  $c_1=M/\left(1-(\epsilon/z_m)^{2-2\gamma_m}\right ) $ and $c_2=\Sigma$. 
In terms of the renormalized quantities 
Eqs.(\ref{vcoffi1}) and (\ref{vcoffi2}) are rewritten as: 
\begin{eqnarray}
M &=& 
{\cal {M}}-\left(\frac{\epsilon}{z_m}\right)^{2-2\gamma_m} \left(\frac{L}{z_m}\right)^{1+\gamma_m}\frac{\xi}{L}
\,, \label{c1:M} \\ 
\Sigma &=& 
 \left(\frac{L}{z_m}\right)^{3-\gamma_m}\frac{\xi}{L}
- 
\frac{
\left(\frac{L}{z_m}\right)^{2-2\gamma_m} 
}{
1-\left(\frac{\epsilon}{z_m} \right)^{2-2\gamma_m} 
} M \, . 
\label{c2:M}
\end{eqnarray}
In the chiral symmetric limit $M = 0$,  we have
\begin{equation}
\label{Sigma:ch}
\Sigma = \left(\frac{L}{z_m}\right)^{3-\gamma_m}\frac{\xi}{L}.
\label{chirallimitcondense}
\end{equation}

 By substituting Eq.(\ref{vsol}) into Eq.(\ref{vaction}), 
the generating functional for $\condense$ is expressed as 
\begin{equation}
\label{W:cond}
W[{\cal M}]=\int d^4x \frac{L}{2g_5^2}\left[ \frac{-L^2}{z^3}\partial_z v(z) \cdot  v(z) \right]^{\zm}_{\ep}.
\end{equation}
From Eq.(\ref{TbarT}) 
we find the techni-fermion condensate $\condense$:
\begin{eqnarray}
\condense
=-\frac{1}{L^3}\frac{L}{g_5^2}(3-\gamma_m)
\left(\frac{L}{z_m}\right)^{3-\gamma_m}\xi \, ,
\label{condenseL}
\end{eqnarray} 
where we have used Eq.(\ref{vcoffi2}) to rewrite $\Sigma$ in terms of $\xi$ and $M$.
From this form, we see that 
the IR value $\xi$ is actually associated with 
the techni-fermion condensate $\condense$ (in a combination with $z_m$, however). 
From Eqs.(\ref{Sigma:ch}) and (\ref{condenseL}) we see that $\Sigma$ is more directly  related to $\condense$ 
(without combination with $z_m$) as 
\begin{eqnarray}
\Sigma =-\frac{ g_5^2/L} {3-\gamma_m} \cdot \condense L^2\,.
\label{Sigmatocondense}
\end{eqnarray}

\subsection{Generating Functional $W[v_\mu,a_\mu]$ and $\Pi_{V,A}$}

 Under the gauge-fixing condition (\ref{gauge-fixing}), 
 we may find the equations of motion for the transversely polarized 
component of the gauge fields $V_\mu(x,z)$ and $A_\mu(x,z)$, 
\begin{eqnarray}
&& 
\left[\partial^2-z\partial_z\frac{1}{z}\partial_z\,\right]V_{\mu}(x,z)=0 
\,, \label{eom:V} \\
&& \left[\partial^2-z\partial_z\frac{1}{z}\partial_z+
\frac{2L^2 v^2(z)}{z^2}\,\right]A_{\mu}(x,z)=0
\, . 
\label{eom:A}
\end{eqnarray}
In solving these equations, 
it is convenient to perform partially Fourier transformation 
on $V_\mu(x,z)$ and $A_\mu(x,z)$ 
with respect to $x_\mu$, 
\begin{eqnarray} 
V_\mu(x,z) = \int_q \, e^{iqx} V_\mu(q,z) \,,\qquad 
A_\mu(x,z) = \int_q \, e^{iqx} A_\mu(q,z) \,, 
\end{eqnarray} 
where the Fourier components $V_\mu(q,z)$ and $A_\mu(q,z)$ 
may be decomposed as 
\begin{eqnarray} 
V_{\mu}(q,z) = \tilde{v}^{\mu}(q)\,V(q,z) \,,\qquad  
A_{\mu}(q,z) = \tilde{a}^{\mu}(q)\,A(q,z) 
\,.
\end{eqnarray} 
Putting these into Eqs.(\ref{eom:V}) and (\ref{eom:A}), 
we have  
\begin{eqnarray}
&&\left[q^2+z\partial_z\frac{1}{z}\partial_z\,\right]V(q,z)\,=\,0
\label{VEOM} , \\
&&\left[q^2+z\partial_z\frac{1}{z}\partial_z-
\frac{2L^2 v^2(z)}{z^2}\,\right]A(q,z)\,=\,0 , 
\label{AEOM}
\end{eqnarray}
with the boundary condition 
\begin{eqnarray}
\label{VAUVBC}
\partial_z V(q,\zm) &=& \partial_z A(q,\zm)\,=\,0, \\
\label{VAIRBC}
V(q,\ep) &=& A(q,\ep)\,=\,1. 
\end{eqnarray}

 The generating functional $W[v_\mu, a_\mu]$ 
is now expressed in terms of $V(q,z)$ and $A(q,z)$ 
as follows: 
\begin{equation}
W[v_\mu,a_\mu] = \frac{1}{2}\int_q \,\frac{-L}{g_5^2\epsilon}
\tr \left[\tilde{v}_{\mu}(-q)\partial_zV(q,\ep)\cdot \tilde v^{\mu}(q)
+\tilde a_{\mu}(-q)\partial_zA(q,\ep)\cdot \tilde a^{\mu}(q)\right]
\,.
\end{equation} 
Accordingly, the vector and axialvector current correlators 
$\Pi_{V}$ and $\Pi_A$, defined as in Eqs.(\ref{Pi:V}) and (\ref{Pi:A}), 
take the form: 
\begin{eqnarray}
\Pi_V(Q^2) = \frac{L}{g_5^2\epsilon}\partial_zV(Q^2,\ep) 
 \,, \label{Pi:Vp} \qquad
\Pi_A(Q^2) = \frac{L}{g_5^2\epsilon}\partial_zA(Q^2,\ep) 
\,, \label{Pi:Ap}
\end{eqnarray}
where we have rewritten $V(q,z) = V(Q^2,z)$ and $A(q,z)=A(Q^2,z)$.

It is now obvious that the chiral symmetry breaking effects  described by  $\Pi_V(Q^2)-\Pi_A(Q^2)$
are related to $\partial_zV(Q^2,\ep) - \partial_zA(Q^2,\ep)  $ and hence  arise  only from the  
$v(z)$ term in Eq.(\ref{eom:A}) which is the unique origin of the $\gamma_m$-dependence in this approach. 
If  the chiral symmetry gets restored $\condense\to 0$ such that $v(z) \to 0$,  we should get 
 $[\Pi_V(Q^2)-\Pi_A(Q^2)] \to 0$, and hence at first glance,  its derivative $\hat S \sim 
\frac{\partial}{\partial Q^2} \left[ \Pi_V(Q^2)-\Pi_A(Q^2) \right] \Big|_{Q=0}$ would also vanish. 
However, overall absolute value of  $\Pi_V(Q^2)-\Pi_A(Q^2)$ is normalized by $f_\pi^2=[\Pi_V(0)-\Pi_A(0)] \to 0$ 
so that   $[\Pi_V(Q^2)-\Pi_A(Q^2)] \to 0$ is realized even with $\hat S \ne 0$. Of course, if we 
have  $[\Pi_V(Q^2)-\Pi_A(Q^2)] \equiv 0$, then $\hat S\equiv 0$ as it should. The situation is very much like the
perturbative calculation of $\hat S$: There could be discontinuity at the phase transition point.
 It should also be noted that although $v(z) \to 0$ in the chiral limit,  $M=0$, implies  $\Sigma \to 0$ and 
$\condense \to 0$, it does not necessarily  $\xi \to 0$. 
This is because that $1/z_m \to 0$ is also possible in the expression of Eq.(\ref{condenseL}). 
The $\hat S$ is given as a function of $\xi$ which is a combination of 
 $\condense$ and $z_m$, so that
its behavior near the conformal phase transition  is not directly connected with $\condense \to 0$.
In the following sections we shall discuss
 these points carefully.

\subsubsection{Vector Current Correlator $\Pi_V$}

A solution of Eq.(\ref{VEOM}) with the boundary conditions 
(\ref{VAUVBC}) and (\ref{VAIRBC}) taken into account 
is given by  the modified Bessel functions $I$ and $K$,
\begin{equation}
V(Q^2,z)=\frac{ K_{0}(Qz_m) \cdot 
       I_{1}(Q z)
       +I_{0}(Qz_m) \cdot 
       K_{1}(Q z) 
       } 
     { I_{0}(Qz_m) \cdot 
       K_{1}(Q\ep) 
      +K_{0}(Qz_m) \cdot 
       I_{1}(Q\ep) }
       \,,
\end{equation}
 and hence $\Pi_V$ in Eq.(\ref{Pi:Vp}) is given by 
\begin{equation}
\label{PiV}
\Pi_V(Q^2)=\frac{L}{g_5^2}\frac{Q}{\epsilon}
\frac{ K_{0}(Qz_m) \cdot 
       I_{0}(Q\ep)
       -I_{0}(Qz_m) \cdot 
       K_{0}(Q\ep) 
       } 
     { I_{0}(Qz_m) \cdot 
       K_{1}(Q\ep) 
      +K_{0}(Qz_m) \cdot 
       I_{1}(Q\ep) }
       \,.
\end{equation}
In particular, for small $Q^2$, 
we may calculate approximately 
\begin{equation} 
 \Pi_V(Q^2\rightarrow 0) \sim 
-\frac{LQ^2}{g_5^2}\log{(\frac{z_m}{\epsilon})}
\,.  
\label{PiVzero}
\end{equation}
We will later come back to this expression  
in evaluating the $S$ parameter.

\subsubsection{Axialvector Current Correlator $\Pi_A$}

  To derive the solution for $\Pi_A$ in an analytic manner~\cite{Hong:2006si}, 
we may define the following quantity: 
\begin{equation} 
\label{Pdef}
P(Q^2,z) = \frac{1}{z}\partial_z \log{A(Q^2,z)}
\, ,
\end{equation} 
in terms of which $\Pi_A$ is expressed as
\begin{equation}
\Pi_A(Q^2) \,=\, \frac{L}{g^2_5 }P(Q^2,\ep).
\label{PiA-P}
\end{equation}
Equation of motion (\ref{AEOM}) is rewritten as 
\begin{equation}\label{TAEOM}
z\partial_zP(Q^2,z)+z^2P(Q^2,z)^2
-Q^2-2\left(\frac{L\,v(z)}{z}\right)^2 =0,
\end{equation}
with the boundary condition 
\begin{equation} 
P(Q^2,\zm)=0 .
\label{BC:P}
\end{equation}

We expand $P(Q^2,z)$ perturbatively in powers of 
 $Q^2$ as 
\begin{equation} 
 P(Q^2,z) = P(0,z)+Q^2P'(0,z)+ \cdots 
\,, \label{P-expand}
\end{equation}
where $  P'(0,z)  \equiv \partial P(Q^2,z)/ \partial Q^2 \Big|_{Q^2=0}$.
These expansion coefficients are determined by solving 
the following equations order by order in $Q^2$: 
\begin{eqnarray}\label{pzero}
O(Q^0) &:& 
z\partial_zP(0,z)+z^2\left(P(0,z)\right)^2 
= \frac{2L^2v^2(z)}{z^2}\,, 
\label{q1} 
\\ 
O(Q^2) &:& 
z\partial_zP'(0,z)+2z^2P(0,z)P'(0,z) 
= 1 \, , \label{q2}
\end{eqnarray}  
which are derived from Eqs.(\ref{TAEOM}) and (\ref{P-expand}). 
Inserting into Eq.(\ref{q1}) the solution of $v(z)$ given in Eq.(\ref{vsol}) 
with $M=0$ taken, 
we may find a solution of Eq.(\ref{q1}) so as to 
satisfy the boundary condition (\ref{BC:P}): 
\begin{eqnarray}
P(0,z)=
\frac{\Delta X(z)}{ z \ep}
\frac{ I_{\frac{1-\Delta}{\Delta}}(X(\zm)) \cdot 
       K_{\frac{1-\Delta}{\Delta}}(X(z)) 
     - K_{\frac{1-\Delta}{\Delta}}(X(\zm)) \cdot 
       I_{\frac{1-\Delta}{\Delta}}(X(z))} 
     { I_{\frac{1-\Delta}{\Delta}}(X(\zm)) \cdot 
       K_{\frac{1}{\Delta}}(X(\ep)) 
     + K_{\frac{1-\Delta}{\Delta}}(X(\zm)) \cdot 
       I_{\frac{1}{\Delta}}(X(\ep))}
\,,
\label{P0z}
\end{eqnarray}
with $\Delta\,=\,3-\gamma_m$ and 
$
X(z) = \frac{\sqrt{2}\xi}
{3-\gamma_m}\left(\frac{z}{\zm}\right)^{3-\gamma_m}
\,. 
$
  Using that solution, we successively solve Eq.(\ref{q2}): 
\begin{equation}\label{P'02}
P'(0,z)=\int_{z}^{z_m}\frac{ d\,z^{\prime}}{z^{\prime}}\,
\left(A(0,z^{\prime})\right)^2\,,
\end{equation}
where  
\begin{equation}
\label{A0}
A(0,z) = 
\frac{z}{\ep}
\frac{ I_{\frac{1-\Delta}{\Delta}}(X(\zm)) \cdot 
       K_{\frac{1}{\Delta}}(X(z)) 
     + K_{\frac{1-\Delta}{\Delta}}(X(\zm)) \cdot 
       I_{\frac{1}{\Delta}}(X(z))} 
     { I_{\frac{1-\Delta}{\Delta}}(X(\zm)) \cdot 
       K_{\frac{1}{\Delta}}(X(\ep)) 
     + K_{\frac{1-\Delta}{\Delta}}(X(\zm)) \cdot 
       I_{\frac{1}{\Delta}}(X(\ep))}
\,. \label{A0}
\end{equation}

Then $\Pi_A(0)$ and $\Pi_A'(0)$ are respectively given by the expansion coefficients $P(0, z)$ and $P'(0,z)$:
\begin{eqnarray}
&\Pi_A(0) & =
\frac{L\,P(0,\ep)}{g_5^2}  \nonumber \\
&=& \frac{L}{g_5^2}\frac{2\Delta}{z_m^2} 
\left(\frac{\xi}{\sqrt 2 \Delta}\right)^{2/\Delta}
\frac{\Gamma(\frac{\Delta-1}{\Delta})}{\Gamma(\frac{1}{\Delta})}
\left(\frac{2}{\pi} \sin \left(\frac{\pi}{\Delta} \right) 
\frac{K_{\frac{1-\Delta}{\Delta}}(X(\zm))}{I_{\frac{1-\Delta}{\Delta}}(X(\zm))} -1\right)
 + {\cal{O}} \left(\left(\frac{\ep}{\zm}\right)^{2\Delta}\right),
\label{PiA0}\\
&\Pi_A'(0) & = 
\frac{L\,P'(0,\ep)}{g_5^2} 
=\frac{L}{g_5^2}\int_{\ep}^{z_m}\frac{ d\,z^{\prime}}{z^{\prime}}\,
\left(A(0,z^{\prime})\right)^2
\,.\label{PiA'0}  
\end{eqnarray}
$\Pi_A(0)$ yields the decay constant $f_\pi$, $\Pi_A(0) = - f_\pi^2$,
while $\Pi_A'(0)$ is related to $\hat S$:
$\hat{S} = -4\pi \left[\Pi_V'(0) - \Pi_A'(0)\right]$.

\section{The $S$ Parameter in Holographic Technicolor}
\label{S-parameter}

Now that we have calculated two-point functions $\Pi_V$ and $\Pi_A$, 
we can compute $\hat{S}$ defined as in Eq.(\ref{Sdif}):
\begin{eqnarray}
\hat{S} 
&=& 
\frac{4\pi L}{g_5^2}\left[ 
\log{\frac{z_m}{\epsilon}} 
   - z \,\int_{\epsilon}^{z_m} \frac{ d z^{\prime}}{z^{\prime}} \,
\left(A^{(0)}(z^{\prime})\right)^2 \right] 
\nonumber \\ 
&=& 
\frac{4 \pi L}{g_5^2} \,\int_{\epsilon}^{z_m}\frac{ d z^{\prime}}{z^{\prime}}\,
\left[1-\left(A^{(0)}(z^{\prime})\right)^2 \right]
\,, \label{S-VA}
\end{eqnarray} 
where use has been made of Eqs.(\ref{PiVzero}) and (\ref{PiA'0}).
We now evaluate $\hat{S}$ 
and show that $\hat{S}$ 
depends only on the ratio $f_\pi/M_\rho$ in the limit $\ep \to 0$, once the value of $\gamma_m$ is fixed, 
where $M_\rho$ and $f_\pi$ denote respectively the mass of techni $\rho$ meson and 
the decay constant.

\subsection{Parameters Relevant to $\hat{S}$}

Let us first recall that 
the original 5-dimensional holographic model analyzed in this paper 
is described by six parameters, 
$(L/g_5^2)$, 
$z_m$, $\epsilon$, $\gamma_m$, $\xi$, and ${\cal {M}}$ (or $M$). 
As far as calculation of the $S$ parameter is concerned, 
it is sufficient to work in the chiral limit $M = 0$, in which case ${\cal{M}}$ is related to $\xi$ 
as seen in Eq.(\ref{c1:M}). 
It should also be noted that
the dimensionful parameters $\ep$ and $z_m$ enter the dimensionless quantity $\hat{S}$
only through the ratio $\ep/z_m$.
In the TC scenario $\ep$ is taken to be the ETC scale $1/\ep = \Lambda_{\rm{ETC}}$ 
and hence $\ep/z_m \ll 1$. 
Here we simply put $\ep/z_m = 0$. 
On the other hand, the decay constant $f_\pi$, $f_\pi^2 = - \Pi_A(0)$, 
depends solely on the dimensionful parameter $z_m$.

The number of parameters relevant to $\hat{S}$ and $f_\pi^2$ hence results in 
three and four, respectively:  
\begin{eqnarray}  
\hat{S} &=& 
\hat{S}
\left( L/g_5^2 ; \, \gamma_m ; \, \xi \right)  
\,, \label{para:S}\\
f_\pi^2 &=& f_\pi^2
\left( L/g_5^2 ; \, \gamma_m ; \, \xi ;\, z_m\right)  
\,. \label{para:fpi} 
\end{eqnarray} 
Although the holography gives us $\hat S$ and $f_\pi^2$ as functions of these parameters
as in Eqs.(\ref{PiV}), (\ref{PiA0}), and (\ref{PiA'0}), 
values of the parameters are not calculable in the framework of the present holographic approach.
In the following,  thereby, we shall discuss how these parameters would behave 
in the framework of walking/conformal TC
with large anomalous dimension $\gamma_m$.

\subsubsection{Parameter $L/g_5^2$}

    The parameter $(L/g_5^2)$ can be determined 
by requiring that the high energy behavior of 
the current correlator $\Pi_V$ should match 
with those of corresponding TC theory in large $N_{TC}$ limit. 
Let us look at the large momentum behavior of $\Pi_V$ derived from Eq.(\ref{PiV}): 
\begin{equation}
\Pi_V(Q^2 \to \infty) 
\sim 
\frac{LQ^2}{2g_5^2}\log{(Q^2\epsilon^2)}
\,. 
\label{PiVinf}
\end{equation}
This expression may be matched with that calculated from the operator product expansion (OPE) 
in the large $N_{TC}$ limit, 
\begin{equation}\label{PiV:OPE}
\Pi_V(Q^2\rightarrow \infty)\sim \frac{N_{TC}Q^2}{24\pi^2}\log{(Q^2\epsilon^2)} 
\, ,
\end{equation}
so that $(L/g_5^2)$ is determined as~\cite{DaRold:2005zs,Erlich:2005qh} 
\begin{equation}
\label{g5}
\frac{L}{g_5^2}=\frac{N_{TC}}{12\pi^2}
\, .\label{match:condi}
\end{equation}

Thus it turns out that once $N_{TC}$ is fixed, $\hat{S}$ and $f_\pi^2$ depend only on 
two and three parameters, respectively: 
\begin{eqnarray} \label{Shol}
  \hat{S} 
&=& \hat{S} (\gamma_m; \xi)  
\nonumber  \\ 
&=& 
\frac{N_{TC}}{3\pi} \,\int_{\epsilon}^{z_m}\frac{ d z^{\prime}}{z^{\prime}}\,
\left[1-\left(A^{(0)}(z^{\prime})\right)^2 \right]   
\,, \label{function:S}\\
f_\pi^2 &=& f_\pi^2
\left(\gamma_m ; \, \xi ;\, z_m\right)  
\,, \label{3para:fpi}
\end{eqnarray} 
where we have used Eq.(\ref{match:condi}). 
Notice from Eq.(\ref{A0}) that the $\xi$- and $\gamma_m$-dependences 
are embedded in the expression of $A^{(0)}$.

\subsubsection{Parameters $\xi$ and $\zm$}

\begin{figure}
\begin{center}
   \includegraphics[keepaspectratio=true,height=45mm]{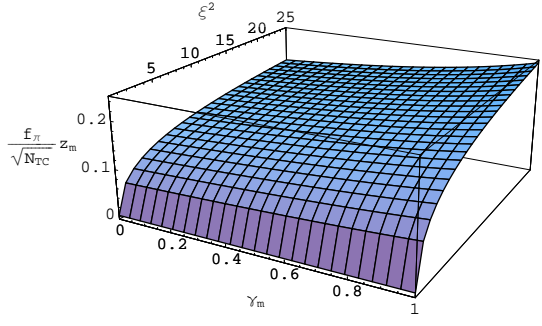}
\end{center}  
  \caption{3-dimensional plot of $\frac{f_\pi}{\sqrt{N_{TC}}} \zm$ 
drawn on ($\gamma_m$, $\xi^2$)-plane.  } 
  \label{fpizm-ad-xi2-3D}
\end{figure}

\begin{figure}
\begin{center}
   \includegraphics[keepaspectratio=true,height=45mm]{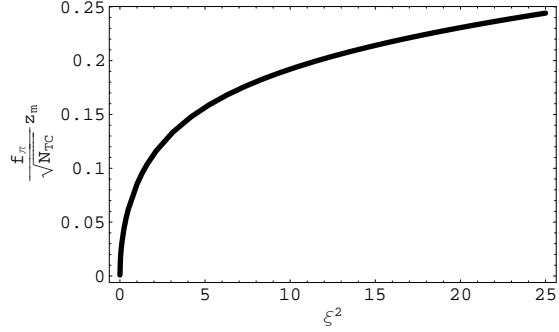}
\end{center}  
  \caption{Plot of $\xi^2$-dependence of $\frac{f_\pi}{\sqrt{N_{TC}}} \zm$ with $\gamma_m \simeq 1$. 
 } 
  \label{fpizm-xi2-ad=1}
\end{figure}

From Eq.(\ref{3para:fpi}) we see that the parameter $\xi$ may be related to $f_\pi$ together with 
the IR brane position $z_m$. 
Equation (\ref{PiA0}) takes the form for $\ep/z_m \to 0$ 
\begin{equation}
f_\pi^2 =\frac{L}{g_5^2}\frac{2\Delta}{z_m^2} 
\left(\frac{\xi}{\sqrt 2 \Delta}\right)^{2/\Delta}
\frac{\Gamma(\frac{\Delta-1}{\Delta})}{\Gamma(\frac{1}{\Delta})}
\left(1-\frac{2}{\pi} \sin \left(\frac{\pi}{\Delta}\right) \frac{K_{\frac{1-\Delta}{\Delta}}(X(\zm))}{I_{\frac{1-\Delta}{\Delta}}(X(\zm))}\right)
 ,
\label{fpi:ep0}
\end{equation}
with $\Delta \equiv 3-\gamma_m$ and $X(z_m) =\sqrt{2}\xi/(3-\gamma_m)$.
 See Figures.~\ref{fpizm-ad-xi2-3D}  and \ref{fpizm-xi2-ad=1}. 
Note that $f_\pi$ is a monotonically increasing function of $\xi$ and $\gamma_m$.

For $\xi \gg 1$ Eq.(\ref{fpi:ep0}) takes the form~\cite{DaRold:2005zs}:
\begin{eqnarray}
f_{\pi}^2
&\stackrel{\xi \gg 1}{\simeq} &
\frac{L}{g_{5}^2}
2^{(1-1/\Delta)}\Delta^{(1-2/\Delta)}
      \frac{ \Gamma(1-1/\Delta)} 
        { \Gamma(1/\Delta)}\frac{\xi^{2/\Delta}}{\zm^2} 
\,.\label{fpi:ximax}
\end{eqnarray}
On the other hand, for $\xi \ll 1$ we have
\begin{eqnarray} 
f_{\pi}^2
&\stackrel{\xi \ll 1}{\simeq} &
\frac{L}{g_{5}^2}\frac{1}{\Delta-1}\frac{\xi^2}{\zm^2} 
\,, \label{fpi:ximin}
\end{eqnarray} 
which coincides with Ref.~\cite{Piai:2006hy} in the case of $\gamma_m = 1$ under certain condition.

The parameter $z_m$ may be related to a typical vector meson mass scale in 
the walking/conformal TC.  
To see this explicitly, we expand 
$\Pi_V$ in terms of the vector meson poles together with the pole resides, 
\begin{equation}\label{NcPiV}
\Pi_V(Q^2)= 
- Q^2\sum_n \frac{F_{V_n}^2}{Q^2+M_{V_n}^2} 
\, ,
\end{equation} 
where $F_{V_n}$ and $M_{V_n}$ denote respectively the vector meson decay constants 
 and their masses. 
Extracting the lightest vector meson-pole, i.e., techni$\rho$-pole, in Eq.(\ref{PiV}) and 
comparing that with Eq.(\ref{NcPiV}), 
we find~\cite{DaRold:2005zs,Erlich:2005qh} 
\begin{equation}\label{rhomass}
M_{V_1} \equiv M_\rho 
\simeq \frac{2.4}{\zm} 
\,. 
\end{equation}

Using Eqs.(\ref{fpi:ximax}), (\ref{fpi:ximin}), and (\ref{rhomass}), we have 
\begin{eqnarray}
\xi^2 &\simeq& \left( \frac{(2.4)^2 C(\gamma_m)}{N_{TC}} \cdot \frac{f_\pi^2}{M_\rho^2} \right)^{3-\gamma_m} 
\,, \qquad 
{\rm for} \, \, \xi \gg 1 
\, , \label{xifpizm:large2}\\ 
\xi^2 &\simeq& \frac{12 \pi^2 (2.4)^2 (2-\gamma_m)}{N_{TC}} \cdot \frac{f_\pi^2}{M^2_\rho} 
\,, \qquad \hspace{5pt}
{\rm for} \, \, \xi \ll 1 
\,, \label{xifpizm:small2} 
\end{eqnarray}
where $C$ is
a numerical factor depending on $\gamma_m$ ($= 3- \Delta$), 
\begin{eqnarray}
C (\gamma_m) &=& 12\pi^2 
2^{(1/\Delta-1)}\Delta^{(2/\Delta-1)}
      \frac{ \Gamma(1/\Delta)} 
        { \Gamma(1-1/\Delta)}.
        \label{coefxiL} 
\end{eqnarray}

\subsection{$\hat{S}$ from Holographic Calculation}

\begin{figure}
\begin{center}
   \includegraphics[keepaspectratio=true,height=45mm]{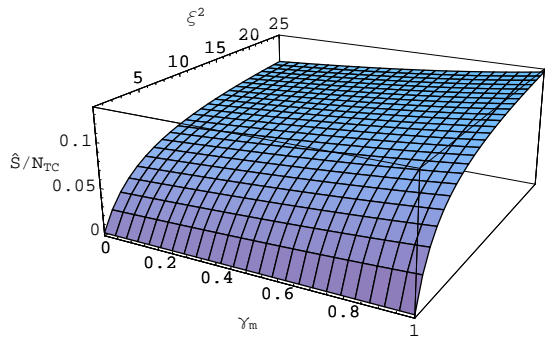}
\end{center}  
  \caption{3-dimensional plot of $\hat S/N_{TC}$ 
drawn on ($\gamma_m$, $\xi^2$)-plane. 
} 
  \label{S-compare}
\end{figure}

\begin{figure}
  \begin{center}
  \includegraphics[keepaspectratio=true,height=45mm]{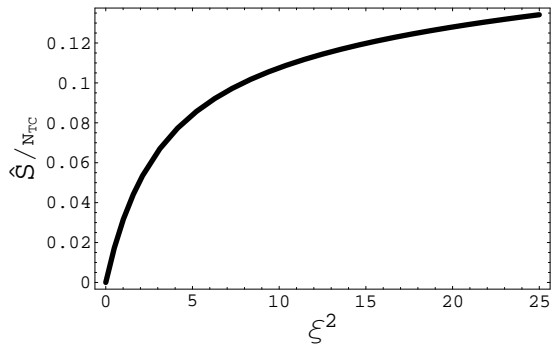}
  \end{center} 
  \caption{Plot of $\xi^2$-dependence of $\hat{S}/N_{TC}$ with $\gamma_m \simeq 1$. 
The blob is the result of the ladder SD and BS equations, $\xi$ from homogeneous 
BS equation~\cite{Harada:2003dc} and $\hat S$ from inhomogeneous BS 
equation~\cite{Harada:2005ru} (to be explained  in Sec.\ref{caseii}).}
  \label{s-xi2-ad=1-ladder}
\end{figure}

We are now ready to evaluate $\hat{S}$ written as a function of $\xi$ 
with large anomalous dimension $\gamma_m$ not restricted to $\gamma_m = 1$. 
Using Eqs.(\ref{S-VA}) and (\ref{match:condi}), we numerically compute $\hat S$, the result given in 
Figs.~\ref{S-compare} and~\ref{s-xi2-ad=1-ladder}. This is our main result.

Varying the values of $\xi$ and $\gamma_m$, 
in Fig.~\ref{S-compare} 
we draw a 3-dimensional plot of ${\hat S}/N_{\rm TC}$ as a function of 
$\xi$ and  $\gamma_m$.   
Also, a plot on (${\hat S}/N_{\rm TC}$, $\xi^2$)-plane for $\gamma_m \simeq 1$ 
is shown in Fig.~\ref{s-xi2-ad=1-ladder}. 
 Looking at Figs.~\ref{S-compare} and~\ref{s-xi2-ad=1-ladder}, 
we find that for smaller value of $\xi (\ll 1)$ $\hat S$ slightly decreases as $\gamma_m$ increases,
while for larger value of $\xi (\gg 1)$ it slightly increases as $\gamma_m$ increases.
Thus there is no dramatic dependence on $\gamma_m$ of $\hat S$ as a function of $\xi$ or $f_\pi/M_\rho$,
although $\gamma_m$ dependence could enter in $\xi$ itself  (we later establish that this is not the case at least explicitly).

In the case of QCD with $\gamma_m \simeq 0$, 
by using phenomenological inputs $f_\pi \simeq 92.4$ MeV and $M_\rho \simeq 775$ MeV,
we estimate $\xi^2 \simeq (4.82)^2$ by Eqs.(\ref{match:condi}), (\ref{fpi:ep0}), and (\ref{rhomass}), 
which in Fig.~\ref{S-compare} implies $\hat S \simeq 0.30$ in agreement with the experiment $\hat S \simeq 0.32$. So the QCD monitor of this holographic approach is checked.
The result is consistent with the estimate of Ref.~\cite{DaRold:2005zs,Hong:2006si}.

Most remarkably, we  find  from Figs.~\ref{S-compare} and~\ref{s-xi2-ad=1-ladder}
that $\hat S$ decreases monotonically with respect to $\xi$, i.e., $f_\pi/M_\rho$, namely
$\hat S \to 0$ can be achieved by taking $\xi \to 0$, or equivalently, $f_\pi/M_\rho \to 0$. 
This tendency is in accord with  Ref.~\cite{Hong:2006si, Piai:2006hy} with $\gamma_m =1$ and also with our later discussion. 
This implies that the holography provides a novel avenue to having 
a small $S$.
Then the next issue is whether or not we can realize $\xi$  in the W/C TC with $\gamma_m \simeq 1$ 
much smaller than $\xi \simeq 4.82$ of QCD with $\gamma_m \simeq 0$~\footnote{
Our result appears to imply that,  
even with small anomalous dimension $\gamma_m \simeq 0$ as in QCD, 
$\hat S$ could be vanishingly small if $f_\pi/M_\rho$ were arranged to vanish.
The point is that in QCD-like theories there is actually no chiral phase transition 
where $f_\pi/M_\rho$ could vanish. 
In contrast, walking/conformal TC characterized by the BZ-IR fixed point does have 
chiral phase transition where $f_\pi/M_\rho$ could have chance to vanish at the phase transition point.
This point will be discussed in details in later section.
}.

Actually, the experimental constraint on $\hat S$ for W/C TC is $S = \frac{N_f}{2} \cdot \hat S < 0.1$. 
In the case of typical W/C TC with $N_{TC} = 2$ and $N_f = 8$ 
we need $\hat S < 0.025$ and hence~\footnote{
In the case of  ``minimal walking''~\cite{Hong:2004td}, the constraint could be  somewhat  weaker.
}
\begin{equation}
\xi^2 < (0.59)^2\,, \,\,{\rm or} \,\,\, \frac{f_\pi}{M_\rho} <  0.038
\label{xi:constraint}
\end{equation}
from Fig.~\ref{s-xi2-ad=1-ladder}, which corresponds to the techni-$\rho$ mass $M_\rho > 3.3 {\rm TeV}$.
In the later section we shall discuss whether or not 
the situation $\xi \ll1$ can be realized in W/C TC with $\gamma_m \simeq 1$.

For later convenience
let us next derive an analytic expression of $\hat S$ for $\xi \gg 1$ and $\xi \ll 1$. 
For $\xi \gg 1$,
from Eq.(\ref{S-VA}) with Eq.(\ref{A0}), we have 
\begin{eqnarray}
\hat{S}
~~\stackrel{\xi \gg 1}{\simeq}~~
\frac{N_{TC}}{3\pi (3-\gamma_m)}
\ln \xi   
\,,
\label{Slargexi}
\end{eqnarray}
which is in accord with the expression obtained in Ref.\cite{DaRold:2005zs}. 
As read off from Eq.(\ref{Slargexi}),  
$\hat S$ cannot be smaller than ${\cal{O}}(1)$ in the case of  $\xi \gg 1$, 
which indicates that this case is phenomenologically unacceptable for TC.

\begin{figure}
\begin{center}
   \includegraphics[keepaspectratio=true,height=45mm]{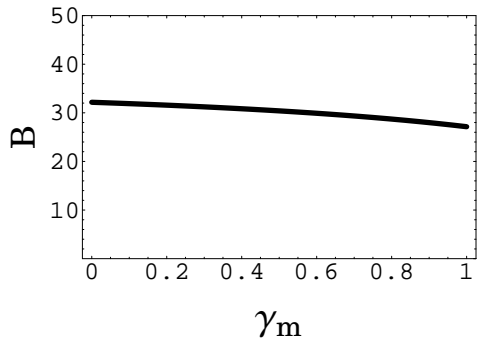}
\end{center}  
  \caption{Plot of the $\gamma_m$-dependence of $B$ (see text).
 } 
  \label{B}
  \label{FigforB}
\end{figure}

 We turn to the case that $\xi \ll 1$:
\begin{equation} 
\hat{S} 
~~\stackrel{\xi \ll 1}{\simeq}~~
\frac{N_{TC} }{6 \pi} \frac{4-\gamma_m}{(3-\gamma_m)^2} \xi^2 
 \, .
\label{Ssmallxi}
\end{equation} 
It is interesting to note that the right hand side of Eq.(\ref{Ssmallxi}) is rewritten 
by using Eq.(\ref{xifpizm:small2}): 
\begin{eqnarray}\label{Sxismall:fpi}
\hat{S}
&\simeq& 
 B \cdot \left(\frac{f_\pi}{M_\rho}\right)^2
\,, \qquad {\rm{for}}\,\, f_\pi/M_\rho \ll 1 ,\label{Ssmallxif} \nonumber \\
B &=& 2\pi (2.4)^2  
 \frac{(2-\gamma_m)(4-\gamma_m)}{(3-\gamma_m)^2},
\label{a-result}
\end{eqnarray}
which shows that  
$\hat{S}$ is given 
as a function of $f_\pi/M_\rho$ 
once $\gamma_m$ is specified.
As already noted in the numerical calculation of $\hat S$,  $B$ in the analytical form is fairly independent of $\gamma_m$
unless $\xi$ or $f_\pi/M_\rho$ is subject to further substantial reduction due to large $\gamma_m$ (we shall discuss this point later):
$B \simeq 27$ for $\gamma_m \simeq 1$ while $B \simeq 32$ as the QCD monitor value  for 
$\gamma_m \simeq 0$ (see Fig.~\ref{B}).

The authors in Ref.~\cite{Hong:2006si} numerically computed $\hat S$ 
as a function of $f_\pi/M_\rho$, focusing only on the case 
with $\gamma_m = 1$ (besides $\gamma_m =0$)~\footnote{The numerical result of Ref.\cite{Hong:2006si} with $C_1 = 0$ (see footnote~\ref{foot:coeff}) 
is somewhat different from ours which corresponds to $C_2 \to 0$ in the limit $\gamma_m \to 1$.
Subtlety about $\gamma_m=1$ will be discussed in Appendix  \ref{cond:gamma1}
}. 
The analytic calculation of $\hat{S}$ was also done in Ref.~\cite{Piai:2006hy} with $\gamma_m = 1$, 
which resulted in the form $\hat{S} \sim (f_\pi/M_\rho)^2$ for $f_\pi/M_\rho \ll 1$.  
Equation (\ref{Sxismall:fpi}) is the analytical expression of our main result.
It implies that $\hat{S} \to 0$ as $f_\pi/M_\rho \to 0$ 
fairly independently of $\gamma_m$.
This suggests existence of a class of phenomenologically viable models of 
walking/conformal TC with $\gamma_m \simeq 1$.

At this point one might suspect that our result (\ref{a-result}) is rather trivial, since we already know 
(see e.g. Ref.~\cite{Harada:2003jx}) the $\rho$-pole-dominated 
expression for $\hat S$ (or $L_{10}=- {\hat S}/(16\pi)$):
\begin{equation}
\hat S \simeq 4 \pi \frac{F_\rho^2}{M_\rho^2} =4 \pi a \frac{f_\pi^2}{M_\rho^2} \, \,\left(
=\left(\frac{g^2}{4\pi}\right)^{-1}\right),
\label{HLS}
\end{equation}
where $F_\rho\, (=\sqrt{a}  f_\pi) $ is the decay constant of  the $\rho$ meson (or, that of the
 NG boson absorbed into longitudinal $\rho$ in the HLS model language),  with $a \simeq 2$ in QCD case by the experiments,
and $g$ the gauge coupling constant of the HLS.
This actually scales as $\hat S \sim f_\pi^2/M_\rho^2$. What is nontrivial with our result in this paper
 is that the holographic calculation 
includes all the  contributions of the poles
not restricted to the lowest one and yet the  coefficient $B$  has no nontrivial dependences on other parameters.
To see this we may write $\hat{S}$ 
in terms of the vector and the axialvector meson pole-dominated expression,  
\begin{eqnarray}
\hat{S} =4\pi \sum_{n=1}^{\infty} \left(\frac{A_{V_n}}{c_{V_n}}-\frac{A_{A_n}}{c_{A_n}}\right)
\cdot \left(\frac{f_\pi}{M_\rho}\right)^2 
\, , \label{Shat:poledomi}
\end{eqnarray}
where 
\begin{eqnarray}
A_{V_n,A_n} \equiv \left( \frac{F_{V_n,A_n} }{ f_\pi } \right)^2, \qquad
c_{V_n,A_n} \equiv \left( \frac{ M_{V_n,A_n} }{M_\rho} \right)^2,
\end{eqnarray} 
with $F_{V_n,A_n}$ and $M_{V_n,A_n}$ being respectively 
the (axial-)vector meson decay constants and (axial-)vector meson masses. 
Note that, 
although each coefficient $A_{V_n,A_n}$ and $c_{V_n,A_n}$ cannot easily be calculated 
without solving non-perturbative issues, 
we may compute a sum of infinite set of pole contributions comparing Eq.(\ref{Ssmallxif}) with Eq.(\ref{Shat:poledomi}):
\begin{eqnarray}
\sum_{n=1}^{\infty} \left(\frac{A_{V_n}}{c_{V_n}}-\frac{A_{A_n}}{c_{A_n}}\right)
= \frac{B}{4\pi}
 \,. \label{holoformula}
\end{eqnarray}

\section{Estimation of $\xi$, or $f_\pi/M_\rho$} 
\label{RGpoint}

In the previous section, we showed that in the holographic calculation 
$\hat S$ is a monotonically increasing function of $\xi$ and  
$\hat{S} \to 0$ as $\xi \to 0$ (Eq.(\ref{Ssmallxi})), fairly independently of $\gamma_m$.
Thus the problem is whether or not the situation $\xi \ll 1$ as in Eq.(\ref{xi:constraint}) can be realized in the holographic W/C TC. 
We may recall that $\xi$ is related to the techni-fermion condensate $\condense$ as in Eq.(\ref{condenseL})
and also to $f_\pi/M_\rho$ as in Eq.(\ref{fpi:ep0}) or Eqs.(\ref{xifpizm:large2}) and (\ref{xifpizm:small2}).
In the following, through correct identification of renormalization-point of $\condense$,
we shall demonstrate that $\xi$ has no particular suppression factor due solely to $\gamma_m$ and so does $S$ 
in contrast to
previous authors~\cite{Hong:2006si, Piai:2006hy}.
Hence the situation $\xi \ll 1$ can only be realized 
near the conformal phase transition point i.e., 
chiral symmetry restoration point at the conformal window where  $\gamma_m \to1$ and vanishing of the dynamical mass $m$
of techni-fermion, $m \to 0$, may be  correlated.
 Actually the  straightforward dynamical calculation based on the ladder SD and BS equations~\cite{Harada:2005ru} shows that this does happen in contrast to the holographic calculation performed here.

\subsection{ $\xi$ and Renormalization of $\condense$}
\label{fpi:mscaling}

\subsubsection{Renormalization of $\condense$}

In order to see whether or not a nontrivial dependence of $f_\pi/M_\rho$ on $\gamma_m$ exists without referring to the
conformal phase transition $m \to 0$, 
we shall make a correct identification of the renormalization-point of $\condense$.
Let us go back to the expression of the current mass of 
techni-fermion $M$ and $\xi$ given in Eqs.(\ref{def:M}) and (\ref{vbdry}). 
In the case $\gamma_m <1$~\footnote{
We shall discuss the limit of $\gamma_m \to 1$ in the Appendix }
 we can safely neglect the second term of  Eq.(\ref{c1:M}) for $\ep \to 0$:
\begin{equation}\label{Mq-ads}
M= {\cal {M}} = 
\left(\frac{1/L}{1/\ep}\right)^{-\gamma_m} 
\cdot \left(\frac{L}{\ep}v(\ep)\right) \,.
\label{massren}
\end{equation}
Given the anomalous dimension $\gamma_m\equiv \partial \ln Z_m(\mu)/\partial \ln \mu$, we obtain the mass renormalization constant $Z_m=  \left(\frac{1/L}{1/\ep}\right)^{\gamma_m} $
by integration  from the cutoff scale
$1/\ep$ down to the infrared scale $1/L$
in a standard way.
Then from Eq. (\ref{massren}) we  see that $M$ is the current mass renormalized at $1/L$ and 
\begin{eqnarray}
M_0 \equiv  Z_m \, M =\left(\frac{1/L}{1/\ep}\right)^{\gamma_m} M= \frac{L}{\ep}v(\ep)
\end{eqnarray}
is the bare mass at the cutoff scale.
We may also introduce a bare condensate:
\begin{eqnarray}
\langle \bar T T\rangle_0 =
 i \frac{ i \delta W[M_0]
}{\delta  M_0} \Bigg|_{M_0=0} 
 =
\left(\frac{1/L}{1/\ep}\right)^{-\gamma_m} \cdot \condense = Z_m ^{-1}\, \condense\, ,
\end{eqnarray}
where $\condense$ is given by Eq.(\ref{condenseL}). Then we have a standard multiplicative renormalization
$M_0\, \langle \bar T T\rangle_0=M \, \langle \bar T T\rangle$.
Since $M \equiv  M_{1/L}$ is the external source for $\condense$,  we conclude that 
$\condense$ in Eq.(\ref{condenseL}) is nothing but the techni-fermion condensate renormalized at the $1/L$, $\condense \equiv \condense_{1/L}$.

   Hence the expression of $\xi$ 
can be written by solving Eq.(\ref{condenseL}) inversely as 
\begin{eqnarray}
\xi &=& -\frac{12\pi^2}{N_{TC}}\frac{z_m^3}{3-\gamma_m} 
\left(\frac{L}{z_m}\right)^{\gamma_m} \condense_{1/L}
\,, 
\label{xi-L}
\end{eqnarray}
where we have used the matching condition of Eq.(\ref{g5}).
We may define the techni-fermion condensate renormalized at $1/z_m$,
$\condense_{1/z_m} \equiv \left(\frac{L}{z_m}\right)^{\gamma_m} \condense_{1/L}$, 
in terms of which we rewrite $\xi$ as
\begin{eqnarray}
\xi  = -\frac{12\pi^2}{N_{TC}}\frac{z_m^3}{3-\gamma_m} 
\condense_{1/z_m}
\,, \label{xi-zm}
\end{eqnarray}
from which 
we readily see that $\xi$ is independent of 
the renormalization scale $L$, and accordingly so is 
$\hat{S}$ 
as it should be. 
This is in sharp contrast to the result of the previous authors~\cite{Hong:2006si,Piai:2006hy}  
which explicitly depends on $L$: $\xi \sim (L/\zm)$ for $\gamma_m = 1$, 
with implicit identification of $\condense_{1/L}$ as $\condense_{1/\zm}$. 
Thus even if we take $L \to \ep$, there is no suppression factor due to $\gamma_m$.

\subsubsection{Relationship Between $f_\pi$, $m$ and $M_\rho$}

The renormalization-point dependence of $\condense$ is further given by \cite{Miransky:vk}
\begin{eqnarray} 
\condense_{1/L} 
&=& \left(\frac{1/L}{m}\right)^{\gamma_m}\condense_m 
\,, \label{scaling-law} \nonumber \\
\condense_m &=&  -\frac{N_{TC}}{4\pi^2} \cdot m^3,
\end{eqnarray}
where the dynamical mass of techni-fermion $m$ may be defined as: 
$m\equiv \Sigma(p^2= -m^2)$ for $i S_{F}^{-1} (p)= p \hspace{-.67em}/ - \Sigma(p^2)$. 
Note that $m \to 0$ at the conformal phase transition point. 
Combining Eqs.(\ref{xi-L}) and (\ref{scaling-law}), 
we find 
\begin{equation}
\xi
 =\frac{3}{3-\gamma_m} 
\left(\frac{m}{\zm^{-1}} \right)^{3-\gamma_m} 
\, .\label{xi}
\end{equation}

  Recalling the relationship between $z_m$ and $M_\rho$ given in 
Eq.(\ref{rhomass}), 
we may further rewrite the right hand side of Eq.(\ref{xi}) as  
\begin{equation}
\xi \simeq 
\frac{3 \cdot (2.4)^{3-\gamma_m}}{3-\gamma_m}
\,\left(\frac{m}{M_{\rho}}\right)^{3-\gamma_m}
 \, .  
 \label{xiMrho}
\end{equation} 
Without knowing  further information that 
$m/M_\rho \to 0$ as $m \to 0$ (and $\gamma_m \to 1$), we see from  
Eq.(\ref{xiMrho}) that $\xi$ has no suppression factor 
due to $\gamma_m$ (it even enhances as $\gamma_m$ increases !).

Incidentally,  Eq.(\ref{Sigma:ch}) is rewritten through Eq.(\ref{xi}) as
\begin{equation}
\Sigma= \left(\frac{L}{z_m}\right)^{3-\gamma_m} \frac{\xi}{L}= \frac{3}{3-\gamma_m} \frac{1}{L} \left(m L\right)^{3-\gamma_m}~~\stackrel{m \to 0}{\longrightarrow}~~ 0\,\, ,
\end{equation}
which implies $v(z)=\Sigma (z/L)^{3-\gamma_m} \to 0$ as $m\to 0$, as it should.  
This should be contrasted with $\xi$ which does not necessarily have a direct tie with
the chiral symmetry restoration $m \to 0$.

Equating right-hand sides of Eq.(\ref{xiMrho}) 
and Eqs.(\ref{xifpizm:large2})-(\ref{xifpizm:small2}), 
we find a relation between $f_\pi$ and $m$
which reads 
\begin{eqnarray} 
\xi \gg 1 &:& 
f_\pi^2/N_{\rm TC}\sim  m^2 
\,, \label{fpi-m:large}\\
\xi \ll 1 &:& 
 f_\pi^2/N_{\rm TC}\sim  m^2 \cdot \left( \frac{m}{M_\rho} \right)^{4-2\gamma_m } 
\,,\label{fpi-m:small}
\end{eqnarray}
near the conformal phase transition point $m \to 0$.
It is worth comparing these relationships with those obtained from 
the Pagels-Stokar formula \cite{Pagels:1979hd}(For details, see Appendix~\ref{PS}).   

Equation (\ref{fpi-m:small}) is a novel scaling when $m/M_\rho \to 0$, 
namely $f_\pi$ vanishes much faster than $m$.
As we shall discuss in the following section,
the holography as it stands does allow for  this possibility,
although we know no explicit calculation to have $m/M_\rho \to 0$ at the conformal phase transition.

\section{Holography versus W/C TC}
\label{critical}

In the previous section, we showed that in the holographic calculation 
$\xi$, or $f_\pi/M_\rho$ has no suppression factor due to $\gamma_m$, 
based on the correct identification of the renormalization-point of $\condense$. 
Actually, the holographic framework as it stands cannot calculate the ratio $f_\pi/M_\rho$  
which is left arbitrary, while it should be a calculable quantity in principle e.g. in the lattice gauge theory. 
In this section, comparing our result with explicit computation of $f_\pi/M_\rho$ in other approaches, 
we shall argue whether or not $f_\pi/M_\rho \ll 1$ can be realized near the conformal window, 
namely $f_\pi/M_\rho \to 0$ for $m \to 0$ in W/C TC 
characterized by the Banks-Zaks infrared fixed point.

\bigskip 

\begin{figure}
  \begin{center}
    \includegraphics[keepaspectratio=true,height=60mm]{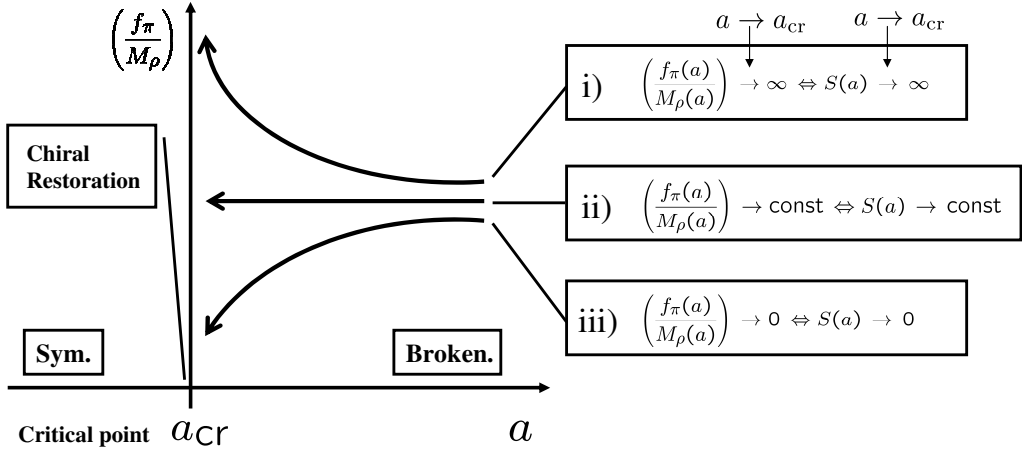}
  \end{center}
  \caption{ An illustration of a classification of holographic walking/conformal technicolor models 
in terms of scaling behaviors of $f_\pi/M_\rho$ near 
the edge of conformal window, which represents $a=a_{cr}$ in this figure, 
where $a$ denotes a tuning parameter characterizing the chiral/conformal phase transition 
by $a=a_{cr}$. 
``Broken." and ``Sym."respectively stands for the broken and the symmetric phases.  } 
  \label{m-Mrho-ratio}
\end{figure}

Without knowledge of detailed dynamics, we may classify the following possibilities 
as illustrated in Fig.~\ref{m-Mrho-ratio} 
on how $f_\pi/M_\rho$ behaves near the conformal phase transition point:

\begin{description}
 
\item {i)} 
$M_\rho \searrow 0$ 
{\it faster than }
$f_\pi \searrow 0$ ($f_\pi/M_\rho \to \infty$). 
In this case 
$\hat{S}$ grows to 
diverge when $m \to 0$.

\item {ii)}
$M_\rho \searrow 0$ 
{\it as fast as} 
$f_\pi \searrow 0$ ($f_\pi/M_\rho \to$constant).
In this case $\hat{S} \to$ constant $\ne 0	$,  
even when $m \to 0$. 

\item {iii)}
$M_\rho \searrow 0$ 
{\it slower} than 
$f_\pi \searrow 0$ ($f_\pi/M_\rho \to 0$).
In this case $\hat{S}$ decreases 
resulting in $\hat{S}=0$ when $m \to 0$.

\end{description}

From Eqs.(\ref{fpi-m:large}) and (\ref{fpi-m:small})
we can read the scaling behavior of $f_\pi$ versus $m$ 
near the conformal phase transition point $m \to 0$ in each case.
In the cases i) and ii) we have a usual scaling $f_\pi/m \to $ constant, while 
in the case iii) where $f_\pi/M_\rho \to 0$ we find a novel scaling behavior of $f_\pi$, $f_\pi/m \to 0$.

\subsection{Searching For Explicit Dynamics}

  Keeping in our mind 
the scaling laws of $f_\pi$, given in Eqs.(\ref{fpi-m:large}) and (\ref{fpi-m:small}), 
in the following we will discuss how the conformal phase transition, classified into 
three types as illustrated in Fig.~\ref{m-Mrho-ratio}, can be realized in explicit W/C TC dynamics.

\subsubsection{Example for  Case i) } 
\label{caseii}

The conformal phase transition corresponding to the case i)   
may be realized by Vector Manifestation (VM)~\cite{Harada:2003jx,Harada:2000kb} in the HLS model.
The $S$ parameter in HLS model is given by Eq.(\ref{HLS}). 
In order that $\hat S <0.1$ (more realistically  $\hat S <0.025$), we would naively need very strong HLS coupling
$\frac{g^2}{4\pi}>10$ (or $\frac{g^2}{4\pi}>40$ !), independently of the tuning of the parameter $a$.
 This is
quite opposite to what is realized in  VM where we have 
 $g\sim \condense_\Lambda/\Lambda^3 \sim (m/\Lambda)^{3-\gamma_m} =(m/\Lambda)^2 \to  0$
as we approach the conformal phase transition point,
$m \to 0$.  Hence we have
\begin{equation}
\hat S = 4\pi a \left(\frac{f_\pi}{M_\rho}\right)^2 =\left(\frac{g^2}{4\pi}\right)^{-1} 
\sim \left(\frac{\Lambda}{m}\right)^4 \to \infty
\,, 
\label{HLSS}
\end{equation} 
where $\Lambda$ is taken as $\Lambda=\Lambda_{\rm TC}=\Lambda_{\rm ETC}$.
This  is similar to the case i) except that the VM yields $\hat S \sim  1/m^4$, while case i) does $\hat S \sim
\ln (f_\pi/M_\rho) \sim \ln (m/M_\rho)$.

\subsubsection{ Example for  Case ii)}

The conformal phase transition, corresponding to the case ii),  
has been indicated by the straightforward calculation of  large $N_f$ QCD
based on the ladder SD equation and the 
BS equation~\cite{Harada:2005ru, Harada:2003dc}. 
   It was shown~\cite{Harada:2003dc} that in the ladder approximation 
homogeneous BS equation together with SD equation (for $N_{\rm TC}=3$ case) gives the bound state mass 
 $M_{\rho}$  as well as $f_{\pi}$, which 
scales  near the conformal window as 
\begin{eqnarray} 
f_\pi 
&\sim &
0.375 
\, m \to 0 \,\,\, ,\,\,\,\,
M_\rho 
 \sim  
4.13 \, m \to 0 
\,, 
\nonumber \\
&&\frac{f_\pi}{M_\rho} \simeq {\rm constant} \simeq 0.091
\,. \label{scaling:2nd}
\end{eqnarray}  
Then we have $(f_\pi  z_m)^2\simeq (2.4)^2 (f_\pi/M_\rho)^2\simeq 0.048$ 
and hence can read off $\xi$ 
from Eq.(\ref{fpi:ep0}) with $\gamma_m\simeq 1$  as (see Fig.~2):
\begin{eqnarray} 
\xi^2 \simeq {\rm constant} \simeq (1.63)^2\, ,
\label{SDBSval}
\end{eqnarray}  
which is $N_{\rm TC}$-independent.

On the other hand, $\hat S$ was straightforwardly calculated through current correlators by the ladder SD equation and inhomogeneous BS equation~\cite{Harada:2005ru}. The result shows that $\hat S$ slightly decreases as we approach
the conformal phase transition point: $\hat S \simeq 0.30$ to $\hat S \simeq 0.25$ (for $N_{\rm TC} =3$):
\begin{eqnarray} 
\hat \frac{S}{N_{\rm TC}} \simeq 0.10 \to 0.083\,\, . 
\end{eqnarray} 
However, since the ladder SD and BS method tends to overestimate $\hat S$ in QCD,
which could be understood as scale ambiguity,   the actual value near the conformal phase transition 
point with $\gamma_m \simeq 1$ 
should be properly re-scaled by a factor roughly 2/3 to fit the QCD value 
correctly when the calculation is extended to the QCD case. If this is done, then the value could be 
\begin{eqnarray} 
 \frac{ {\hat S}^{\rm (re-scaled)}}{N_{\rm TC}}  \simeq 0.067 \to 0.056\,\,. 
  \label{re-scaled}
\end{eqnarray} 

Curiously enough, a set of the values of $\xi^2$ in Eq.(\ref{SDBSval}) and $\hat S/N_{{\rm TC}}$ in Eq.(\ref{re-scaled}) 
fit in  the line of the holographic result in Fig.\ref{s-xi2-ad=1-ladder}. At this moment it is not clear whether or not this coincidence 
has deeper implications.

\subsubsection{Example for  Case iii)}

 As for case iii), at this moment 
we have no  concrete dynamics
which gives rise to the situation that, near the conformal phase transition point $m \to0$,  we have 
$\hat{S}\sim \xi^2  \to 0$ in  such a way that 
\begin{equation}
\xi^2 N_{\rm TC}\sim \left(f_\pi z_m \right)^2\sim\left(\frac{f_\pi}{M_\rho}\right)^2 \to 0 
\,, 
\qquad 
{\rm as} 
\quad 
m \to 0
\,.  \label{scaling:3rd}
\end{equation}
If it is really realized,  our holographic result would imply
 somewhat severe constraint on the value of
$\xi$ in Eq.(\ref{xi:constraint}): $\xi <0.59$ for a typical W/C TC with one family techni-fermions ($N_f=8$) and 
 $N_{\rm TC}$, which actually corresponds to $M_\rho> 3.3{\rm TeV}$. 
Although this might confront 
the perturbative unitarity problem~\footnote{We thank R.S. Chivukula for useful comments on this point.},
this could be resolved by  the techni-dilaton (as a composite Higgs) dynamically formed in the generic W/C TC,
 which could be identified with the bulk scalar in the present holographic approach.

Such a case may be realized in a bizarre situation that in contrast to the usual picture $m\sim z_m^{-1} \sim M_\rho$, 
there may be  no bound states as a remnant of the conformal window 
except for the NG boson  $\pi$ and the techni-dilaton which could be the only light spectra
reflecting the spontaneous chiral symmetry and conformal  symmetry and hence could have arbitrarily small mass
compared with our infrared scale $m \ll z_m^{-1}$ near the
conformal phase transition point.  Although the holography gives an infinite set of bound states consistently with the large $N_c$ limit, the conformal phase transition essentially depends on the Banks-Zaks infrared fixed point which is realized 
only for large $N_f$ (with $N_f/N_c=$ fixed) instead of the large $N_c$ limit.  Thus  all the massive bound states 
would quickly decay into the  constituents i.e. pairs of the light techni-fermions (or, $\pi$'s) 
through the $N_c$ subleading effects.  
We should note that 
such a picture is compared with the  explicit computation based on the ladder SD and BS equations~\cite{Harada:2003dc}  
which actually produce light bound states of $\rho$ and $a_1$ but  $M_\rho, M_{a_1} > 2 m$ in contrast to
 $\pi$ (massless)  and the scalar meson (``techni-dilaton''
with mass $M_{\rm TD} \simeq \sqrt{2} m < 2 m$) near the conformal phase transition point.
Then the vector and axialvector bound states may in principle 
quickly decay into pair of the  light techni-fermions (or  light composite 
$pi$'s).
We will see a definite answer to this possibility by the lattice calculations of  the large $N_f$ QCD in near future.

Another possibility to have small $\hat S$ would be to include  subleading corrections in $1/N_c$ 
to the holography which is valid only at leading order of $1/N_c$.  
Recall that $\hat S$ is given in terms of vector and axialvector pole contributions as in Eq.(\ref{Shat:VD}) .
Although the holography includes all the resonance contributions, it is only the result at $1/N_c$ leading order.
The subleading effects may change the result drastically particularly near the conformal phase transition point where
$N_f$ is large  with $N_f/N_c =$ fixed and we are  considering  corrections to vanishingly small quantities. We can always integrate out higher resonances in  the holographic theory to arrive at the (generalized) 
HLS model which has only few lowest vector and axialvector resonances (``holographic reduction")~\cite{Harada:2006di}.
Then the loop contributions of
this theory  yield part of  the subleading corrections in $1/N_c$ to the holography~\cite{Harada:2006di}. 
Now look at the generalized HLS model containing only  $\rho$ and $a_1$~\cite{Bando:1985rf} as a consequence of the above
 holographic reduction, which corresponds to taking only 
contributions from the lowest resonances  $\rho$ and $a_1$:
\begin{eqnarray}
\hat{S} =4\pi  
\left(
\left(
\frac{F_\rho}{M_\rho}
\right)^2 
-\left(
\frac{F_{a_1}}{M_{a_1}}
\right)^2 
\right) 
= \left(\frac{g^2}{4\pi}\right)^{-1}\left(
 1-
\left(
\frac{b}{b+c} \right)^2  
\right)
\, , 
\label{Shat:VD}
\end{eqnarray}
where $b,c$ are the parameters of this generalized HLS model to be running at loop level, which is compared with the simplest HLS model, Eq.(\ref{HLSS}). 
Differing from the VM based on the one-loop calculations of the simplest HLS model having only
$\rho$ without $a_1$~\cite{Harada:2003jx,Harada:2000kb},  the chiral
restoration due to the one loop contributions of this model is more involved~\cite{Harada:2005br}, 
which may suggest  a possibility of a fixed point of the HLS parameters for giving a vanishing $\hat S \sim (f_\pi/M_\rho)^2 
\sim c/g^2 \to 0$ due to cancellation among $\rho$ and $a_1$ contributions at the chiral restoration point.

Finally, we shall emphasize that, in addition to the scaling relation of $\hat{S}$, 
holographic calculation has provided us with another scaling relation, Eq.(\ref{fpi-m:small}), 
that is the scaling relation of $f_\pi$ with respect to $m$. 
Note that Eq.(\ref{fpi-m:small}) takes a quite different form compared to Eq.(\ref{fpi-m:large}) (which is 
the familiar form, $f_\pi^2/N_f \sim m^2$, as seen in QCD even), 
and hence it could be a key ingredient in a future search for an example of the case iii) 
a development of lattice calculation for large $N_f$ QCD would 
clarify whether such a scaling property can actually be realized.

\section{Summary }
\label{summary}

  In this paper
we have studied   the $S$ parameter in the 
walking/conformal technicolor (W/C TC), based on the deformation of holographic QCD by 
varying the anomalous dimension of techni-fermion condensation from the QCD monitor value $\gamma_m \simeq 0$ to 
that of the W/C TC $\gamma_{m} \simeq 1$. In contrast to the previous authors who worked on $\gamma_m =1$ and particular values of $\xi$, 
we gave an explicit functional form of $\hat{S}$ in the entire parameter space $0<\xi<\infty$ and $0<\gamma_m<1$, which 
turned out to be fairly independent of
the value of $\gamma_m$ and to
behave as
$\hat S \sim \xi^2 N_{\rm TC} \sim (f_\pi/M_\rho)^2$ for $\xi  \ll 1$.
Thus $\hat S \ll 1 $ can be realized,  if we have a dynamics 
showing $\xi^2 N_{\rm TC} \sim (f_\pi/M_\rho)^2 \ll 1 $ near the conformal window where 
the chiral symmetry get restored $\condense \to 0$.  However, although $\xi$ is proportional to $\condense$,
we found no suppression of $\xi$ or $f_\pi/M_\rho$ and hence of  $\hat S$  due solely to the large anomalous dimension
in contrast to the claim of the literature, 
through careful analysis of the renormalization-point dependence of the $\condense$.

Although the ratio $f_\pi/M_\rho$ cannot be calculated in the holography, 
we discussed possible behavior of  it near the conformal window where the chiral symmetry gets restored
$\condense \to 0$.
 To compare the holographic result to that of more explicit dynamics, we classified holographic W/C TC models  
into three cases: 
  i) $f_\pi/M_\rho \to \infty$,  
  ii) $f_\pi/M_\rho \to {\rm constant} \ne 0$, 
  iii) $f_\pi/M_\rho \to 0$.

Case i) roughly corresponds to the Vector Manifestation (VM)~\cite{Harada:2003jx} realized in the simplest HLS model  with $\rho$ and $\pi$ , which yields $\hat S \to \infty$.

Case ii) corresponds to 
the result of the ladder SD and BS equation~\cite{Harada:2005ru,Harada:2000kb} , which to our surprise yields 
not only  $\xi^2 N_{\rm TC}\sim (f_\pi/ M_\rho)^2 \to$ constant but also  a  set of 
the calculated values of $f_\pi/M_\rho$ and $\hat S$ 
well fit to the line of the parameter space of  the holographic result in this paper. 
Deeper implications of this coincidence  
are not clear at this moment.

Although  Case iii) has no explicit dynamics at the moment,   if it is realized, the holographic result we obtained 
 seems to pose a severe constraint on the lower bound of  techni-$\rho$ mass. 
We discussed a possibility for having such a case where there are 
no bound states as a remnant of the conformal window except for the NG boson $\pi$ and the scalar (as a techni-dilaton),   
and hence $m\ll z_m^{-1}$. Actually, the dynamics near the
 conformal phase transition is governed by the Banks-Zaks infrared fixed point due essentially to the large $N_f$ 
 with $N_f/N_c=$ fixed but not to the simple  large $N_c$ limit.
We also discussed another possibility for having $\hat S \,\, \ll 1$ by introducing  $1/N_c$ subleading
 corrections through the meson loops in the generalized HLS model. Note that the generalized HLS model 
 is obtained by the integrating out the higher resonances  of the holographic result which is valid only at $1/N_c$ leading order. 
We also found that 
if the case iii) is realized in some concrete dynamics,  we have a novel scaling property of 
$f_\pi$ with respect to $m$ (Eq.(\ref{fpi-m:small})), 
which takes a quite different form compared to the familiar  form, $f_{\pi}^2/N_{\rm TC}  \sim m^2$. 
This would suggest that this scaling property may play an important role 
to reveal such a phenomenologically viable W/C TC. 

For all these unsolved features, the holographic relation we obtained in this paper would be useful for further
studies of the W/C TC.

\section*{Acknowledgments}

We would like to thank R.S. Chivukula, H. Fukano, M. Harada, D. K. Hong, M. Piai, F. Sannino, S. Sugimoto and H. U. Yee  
for useful comments and fruitful discussions. 
This work is supported in part by the JSPS
Grant-in-Aid for Scientific Research (B) 18340059, The Mitsubishi Foundation,
and Daiko Foundation. K.H thanks COE program at Nagoya University for support of
a trip to Michigan State University where part of this work was done. 
S.M. is supported by the U.S. Department of Energy under
Grants No. DE-FG02-06ER41418. 
K.Y. thanks the Institute for Nuclear Theory
at the University of Washington for its hospitality 
and the Department of Energy for partial support during the completion 
of this work.

\appendix 
\renewcommand\theequation{\Alph{section}.\arabic{equation}}

\section{Limit of $\gamma_m \to1$}
\label{cond:gamma1}
We shall discuss in this Appendix on the limit $\gamma_m \to 1$ of our result 
which we show can be continuously moved over to $\gamma_m=1$.

Let us begin with the classical solution of $v(z)$ of Eq.(\ref{vEOM}) 
for $\gamma_m= 1$, which is given by 
\beq
v(z) ^{(\epsilon)}= C_1 \left(\frac{z}{L}\right)^2 + C_2 \left(\frac{z}{L}\right)^2 \ln{\frac{z}{L}}.
\label{vsolgammam1}
\eeq
On the other hand,  Eq.(\ref{vsol})  for $\gamma_m <1$ takes the form in the limit
 $\gamma_m \to 1$:
\begin{eqnarray}\label{vsol1limit}
v(z)^{(\epsilon)} &=&\left(\frac{z}{L}\right)^2 
\left(c_1 \left(\frac{z}{L}\right)^{-\delta}+c_2 \left(\frac{z}{L}\right)^{\delta}\right) \nonumber \\
&
~~\stackrel{\delta \ll 1}{\simeq}~~ 
&\left(\frac{z}{L}\right)^2 
\left((c_1 +c_2) +(c_2 -c_1)\delta \cdot \ln \left(\frac{z}{L}\right)\right)\, ,
\end{eqnarray} 
where $\delta \equiv 1-\gamma_m$.

Comparing  Eqs.(\ref{vsolgammam1}) and (\ref{vsol1limit}), we read $C_1$ and $C_2$
as
\beqs
C_1& =&\lim_{\delta \to 0}  \left( c_1 + c_2\right) 
=
\lim_{\delta \to 0}  \left[
\left(\frac{L}{z_m}\right)^2 
\frac{\xi}{L} 
+\frac{
\left( 1- \left(\frac{L}{z_m}\right)^{2-2\gamma_m}\right)}{
\left( 1- \left(\frac{\epsilon}{z_m}\right)^{2-2\gamma_m}\right)
}
 \left( {\cal M} -\left(\frac{L}{z_m}\right)^2 \frac{\xi}{L}\right)
\right]
  \nonumber \\
&
=
& 
\left(\frac{L}{z_m}\right)^2 
\frac{\xi}{L} 
+ \frac{\ln \zm/L}{\ln \zm/\ep} \left( {\cal M} -\left(\frac{L}{z_m}\right)^2 \frac{\xi}{L}\right)   
\, ,\\ 
C_2 &=&\lim_{\delta \to 0}  \left[ (c_2- c_1)\delta\right] =  \lim_{\delta \to 0} \left[
\left(
\left(\frac{L}{z_m}\right)^{2} \frac{\xi}{L} - 2\frac{ \left( {\cal M} -\left(\frac{L}{z_m}\right)^2 \frac{\xi}{L}\right)
}{ 1- \left(\frac{\epsilon}{z_m}\right)^{2-2\gamma_m}} 
\right) \delta \right]
\nonumber  \\
&=&
-\frac{1}{
\ln{\frac{\zm}{\ep}}
} \left( {\cal M} -\left(\frac{L}{z_m}\right)^2 \frac{\xi}{L}\right) \, ,
\eeqs
We then find  
\beqs
v(z) ^{(\epsilon)}= C_1 \left(\frac{z}{L}\right)^2 + C_2 \left(\frac{z}{L}\right)^2 \ln{\frac{z}{L}}
= \left(\frac{z}{L}\right)^2
\left[
\frac{\ln \frac{z}{\ep}}{\ln \frac{z_m}{\ep} } \left( \frac{L}{z_m} \right)^{2} \frac{\xi}{L} 
+\frac{\ln \frac{z}{z_m}}{\ln \frac{\ep}{z_m}} {\cal M}
\right] \, ,
%
\eeqs
which obviously satisfies the boundary conditions $v(\ep) ^{(\ep)}= \left(\frac{\ep}{L}\right)^2 {\cal M}$
and $v(z_m) ^{(\ep)}= \frac{\xi}{L}$ as it should. 
Now we may define the current mass $M$ as
\beqs
M \equiv \left[
\left(\frac{L}{\ep}\right)^2
\frac{1}
{\ln \left(\frac{z_m^2}{\ep^2}\right)}
\right] v(\ep) ^{(\ep)}
={\cal M}\left[
\frac{1}
{\ln \left(\frac{z_m^2}{\ep^2}\right)}
\right]
\, ,
\eeqs
which yields
\beqs
v(z) ^{(\epsilon)} &=& 
 \left(\frac{z}{L}\right)^2
\left[
\frac{\ln \frac{z}{\ep}}{\ln \frac{z_m}{\ep} } \left( \frac{L}{z_m} \right)^{2} \frac{\xi}{L} 
- 2 M \ln \left(\frac{z}{z_m}\right) 
\right]
\nonumber  \\
&=& 
\left(\frac{z}{L}\right)^2
\left[
\left(
\frac{\ln \frac{z}{\ep}}{\ln \frac{z_m}{\ep} } 
\left( \frac{L}{z_m} \right)^{2} 
\frac{\xi}{L}  + M \ln \left( \frac{z_m^2}{L^2} \right)
\right)
-2M \ln \frac{z}{L} 
 \right] 
 \, .
\eeqs
When we take the limit $\ep (<z)  \to 0$, we have
\beqs
v(z) ^{(\ep)}~~\stackrel{\epsilon \to 0}{\longrightarrow}~~  v(z) 
=
\left(\frac{z}{L}\right)^2
\left[
\left(
\left( \frac{L}{z_m} \right)^{2} 
\frac{\xi}{L}  + M \ln \left( \frac{z_m^2}{L^2} \right) 
\right)
-2M \ln \frac{z}{L} \right] 
\, .
\eeqs
This takes the form in the chiral limit  $M=0$:
\beqs
v(z) 
=
\left[
\left( \frac{L}{z_m} \right)^{2} 
\frac{\xi}{L} 
 \right]  \cdot\left(\frac{z}{L}\right)^2
\, ,
\eeqs
which is in accord with the $\gamma_m<1$ case, Eqs.(\ref{vsol}) and (\ref{chirallimitcondense}) for $M=0$.
 
 The technifermion condensate is given by  Eq.(\ref{TbarT}) as 
 \beqs
 \condense =
\left(\frac{z_m}{L}\right) \cdot \left[
 -\frac{2}{z_m^3}\frac{L}{g_5^2}\xi\right]
 =Z_m^{-1}(x)\Bigg|_{x=\frac{1/L}{1/z_m}} \cdot \condense_{1/z_m}\, ,
 \eeqs
 where
\beqs
Z_m^{-1} (x)  = x\,,\quad
\condense_{1/z_m}=- \frac{2}{z_m^3}\frac{L}{g_5^2}\xi =  - \frac{N_{TC}}{6\pi^2}  \frac{1}{z_m^3}\xi\, .
\eeqs
Then it is clear that  $\condense$ is the quantity  renormalized  at $1/L$:   $\condense=\condense_{1/L}$
and that $\xi$ does not depend on $1/L$ which corresponds to the renormalization point.
We may introduce the bare mass and bare condensate as
\beqs
\condense_0&=& \condense_{1/\ep}=Z_m^{-1} (L/\ep) \cdot \condense=Z_m^{-1} (z_m/\ep) \cdot \condense_{1/z_m}\, ,
\nonumber \\
M_0&=&Z_m (L/\ep) \cdot M\,, 
\eeqs
so that the multiplicative renormalization of mass operator is evident:
\beqs
M_0 \condense_0= M  \condense \,.
\eeqs
The anomalous dimension is thus given as
\beqs
\gamma_m = \frac{\partial \ln Z_m^{-1}(x)}{\partial \ln x}\Bigg|_{x=L/\ep} =1 \,
\eeqs
in agreement with our procedure in the text taking the limit $\gamma_m (<1)  \to 1$.
 
\section{Relationship Between $f_\pi$ and $m$ in Holography}
\label{PS}

In this section, 
we compare the holographic expression of $f_\pi^2$ 
in terms of $m$ as given in Eqs.(\ref{fpi-m:large}) and (\ref{fpi-m:small}) 
with the Pagels-Stokar formula \cite{Pagels:1979hd} for $f_\pi^2$.

 The Pagels-Stokar formula \cite{Pagels:1979hd} relates 
 the pion decay constant $f_\pi$ to 
 a mass function of techni-fermion $\Sigma(x)$ with $x = - p^2$ as 
\begin{equation} 
 \frac{ (4\pi^2)}{N_c} \cdot f_\pi^2 = 
\int_{\rm IR}^{\rm UV} x dx 
\frac{\Sigma(x)(\Sigma(x)- x \Sigma'(x)/2)}{(x + \Sigma^2(x))^2} 
\,, \label{PSformula}
\end{equation}
where we have introduced ``IR" and ``UV" cutoffs in integral with respect to $x$. 
 It should be noted that, for regions (i) $x> m^2$ and (ii) $x<m^2$, 
the mass function $\Sigma(x)$  can be expressed  
in terms of the dynamical fermion mass $m\equiv \Sigma(m^2)$ with the anomalous dimension $\gamma_m$ 
($0 < \gamma_m < 1$) as 
\begin{eqnarray}
 ({\rm i})\,  x> m^2, &&\quad   \Sigma_{(i)}(x)\sim \frac{m^3}{x}\left(\frac{x}{m^2}\right)^{\gamma_m/2}
\,, \label{momapp1} \\
  ({\rm ii}) \,  x< m^2, &&\quad  \Sigma_{(ii)}(x)\sim m 
  \,.  \label{momapp2}
\end{eqnarray}

To make contact with holographic calculations, 
we may identify IR and UV scales as 
\begin{equation}
  {\rm UV} \equiv (\epsilon^{-1})^2 = \Lambda^2 \,, 
\qquad 
 {\rm IR} \equiv (z_m^{-1})^2 \simeq M_{\rho}^2 
 \,. 
\end{equation}
Note that the integration in $x$ necessarily results in convergence as for $0 < \gamma_m < 1$. 
Therefore we hereafter evaluate $f_\pi$ taking the continuum limit $\Lambda \to \infty$, 
but keep dependence of the IR scale $M_\rho$ in the expression of $f_\pi$.

Let us first examine the case that 
$m > M_\rho$. 
Putting the asymptotic expression for $\Sigma(x)$ given in Eqs.(\ref{momapp1}) and (\ref{momapp2}) into Eq.(\ref{PSformula}), 
we straightforwardly calculate the right hand side of Eq.(\ref{PSformula})  as 
\begin{eqnarray}
   \frac{(4\pi^2)}{N_c} \cdot f_\pi^2(m > M_\rho) 
&=&
\int_{M_\rho^2}^{m^2} x dx 
\frac{\Sigma_{(ii)}(x)(\Sigma_{(ii)}(x)- x \Sigma_{(ii)}'(x)/2)}{(x + \Sigma_{(ii)}^2(x))^2} 
\nonumber \\ 
&& 
+ 
\int_{m^2}^{\infty} x dx 
\frac{\Sigma_{(i)}(x)(\Sigma_{(i)}(x)- x \Sigma_{(i)}'(x)/2)}{(x + \Sigma_{(i)}^2(x))^2}  
\nonumber \\ 
&=&  
\int_{M_\rho^2}^{m^2} x dx 
\frac{m^2}{(x + m^2)^2} 
\nonumber \\ 
&& 
+
 \int_{m^2}^{\infty} x dx 
\frac{\Sigma_{(i)}(x)(\Sigma_{(i)}(x)- x \Sigma_{(i)}'(x)/2)}{(x + \Sigma_{(i)}^2(x))^2} 
 \nonumber \\
&=& 
 m^2\Bigg[ 
\ln{2}-\frac{1}{2}~+~
\left(\frac{3+\Delta}{8\Delta^2}\right)
 \left(\Delta-\psi (\frac{1}{2}-\frac{1}{2\Delta})+\psi (1-\frac{1}{2\Delta})\right) 
\Bigg]
\nonumber \\ 
&&
+
 {\cal{O}}\left(\Big(\frac{M_\rho}{m}\Big)^4\right)
\,,
 \label{fpi-m-M:large}
 \end{eqnarray} 
where 
$\psi$ denotes a poli-gamma function. 
 From Eq.(\ref{fpi-m-M:large}), 
we see that $f_\pi$ scales as $m \to 0$ 
independently of IR cutoff scale $M_\rho$ for any value of  $\gamma_m$, 
\begin{equation}
f^2_\pi (m >  M_\rho) 
\sim m^2
\,. \label{scalePSla}
\end{equation}
which results in the same form as that of Eq.(\ref{fpi-m:large}).

 We next turn to the case $m \ll M_\rho$.  
By putting Eq.(\ref{momapp1}) into Eq.(\ref{PSformula}), 
$f_{\pi}$ may be calculated to be 
\begin{eqnarray}
   \frac{(4\pi^2)}{N_c} \cdot f_\pi^2(m \ll M_\rho) 
&=&  
\int_{M_{\rho}^2}^{\infty} x dx 
\frac{\Sigma_{(i)}(x)(\Sigma_{(i)}(x)- x \Sigma_{(i)}'(x)/2)}{(x + \Sigma_{(i)}^2(x))^2} 
\nonumber \\ 
&=& 
 \frac{m^2}{2-\gamma_m} \cdot \left(\frac{m}{M_\rho}\right)^{2\gamma_m-4}
\nonumber \\ 
&& 
\times 
  F\left[ 
2,\frac{2-\gamma_m}{3-\gamma_m},  
\frac{5-2\gamma_m}{3-\gamma_m}, 
-\left(\frac{m}{M_\rho}\right)^{6-2\gamma_m} 
\right] 
\,, \nonumber \\ 
\label{fpi-m-M:small}
\end{eqnarray} 
where $F$ denote a hyper-geometric function. 
  From Eq.(\ref{fpi-m-M:small}), in the limit $m/M_\rho \to 0$, 
we find a scaling relation sensitive to  both $M_\rho$ and $m$,  
\begin{equation}\label{scalePSsm}
f_{\pi}^2(m \ll M_\rho)
\sim m^2\cdot \left(\frac{m}{M_\rho}\right)^{4-2\gamma_m}
\,, 
\end{equation} 
which is in accord with that of Eq.(\ref{fpi-m:small}).

Thus it turn out from Eqs.(\ref{scalePSla}) and (\ref{scalePSsm}) 
that the scaling relations (\ref{fpi-m:large}) and (\ref{fpi-m:small}) between $f_\pi$ and $m$, 
calculated in holographic technicolors, are exactly reproduced by the PS formula.


\end{document}